\documentclass[jcappub,11pt]{article}
\pdfoutput=1 
\usepackage{jcappub} 
\usepackage[T1]{fontenc} 
\usepackage{url}
\usepackage{xcolor}

\title{\boldmath Probing massive neutrinos with the Minkowski functionals of large-scale structure}

\author[a,b]{Wei Liu,}
\author[a,b]{Aoxiang Jiang,}
\author[a,b,1]{and Wenjuan Fang \note{Corresponding author.}}

\affiliation[a]{CAS Key Laboratory for Research in Galaxies and Cosmology, Department of Astronomy, University of Science and Technology of China, Hefei, Anhui, 230026, P.R.China}
\affiliation[b]{School of Astronomy and Space Sciences, University of Science and Technology of China, Hefei, Anhui, 230026, P.R.China}

\emailAdd{lw980228@mail.ustc.edu.cn}
\emailAdd{jax9709@mail.ustc.edu.cn}
\emailAdd{wjfang@ustc.edu.cn}

\abstract{Massive neutrinos suppress the growth of structure under their free-streaming scales. The effect is most prominent on small scales where the widely-used two-point statistics can no longer capture the full information. In this work, we study the signatures massive neutrinos leave on large-scale structure (LSS) as revealed by its morphological properties, which are fully described by $4$ Minkowski functionals (MFs), and quantify the constraints on the summed neutrino mass $M_{\nu}$ from the MFs, by using publicly available N-body simulations. We find the MFs provide important complementary information, and give tighter constraints on $M_{\nu}$ than the power spectrum. Specifically, depending on whether massive neutrinos are included in the density field (the `m' field) or not (the `cb' field), we find the constraint on $M_{\nu}$ from the MFs with a smoothing scale of $R_G=5 h^{-1}$Mpc is $48$ or $4$ times better than that from the power spectrum. When the MFs are combined with the power spectrum, they can improve the constraint on $M_{\nu}$ from the latter by a factor of  63 for the `m' field and 5 for the  `cb' field. Notably, when the `m' field is used, the constraint on $M_{\nu}$ from the MFs can reach $0.0177$eV with a volume of $1(h^{-1}\rm Gpc)^3$, while the combination of the MFs and power spectrum can tighten this constraint to be $0.0133$eV, a $4.5\sigma$ significance on detecting the minimum sum of the neutrino masses. For the `m' field, we also find the $\sigma_8$ and $M_{\nu}$ degeneracy is broken with the MFs, leading to stronger constraints on all 6 cosmological parameters considered in this work than the power spectrum.}

\begin{document}
\maketitle
\flushbottom

\section{Introduction}
Neutrino oscillation experiments have firmly established the existence of neutrino masses \cite{PhysRevLett.81.1562,PhysRevLett.89.011301,PhysRevLett.94.081801,PhysRevLett.101.131802}. The measured mass splittings, $\Delta m_{21}^{2} \equiv m_{2}^{2}-m_{1}^{2} \approx 7.49_{-0.17}^{+0.19} \times 10^{-5} \mathrm{eV}^{2}$ and  $\left|\Delta m_{31}^{2}\right| \equiv\left|m_{3}^{2}-m_{1}^{2}\right| \approx$ $2.484_{-0.048}^{+0.045} \times 10^{-3} \mathrm{eV}^{2}(1 \sigma)$ \cite{hierarchy},  imply that at least two of the mass eigenvalues are nonzero. Depending on the sign of $\Delta m_{31}^{2}$, there are two possible rankings for the three neutrino masses. Assuming the lightest mass to be zero, the minimum summed mass $M_{\nu}$ for the two rankings are approximately 0.06 eV (normal hierarchy with $m_1<m_2<m_3$) and 0.1eV (inverted hierarchy with $m_2>m_1>m_3$) \cite{Hannestad_2016}. However, to determine the mass ranking and the absolute mass scale is generally a challening task for laboratory experiments \cite{drexlin+13, Salas+18}.

Through a completely independent approach, cosmology offers a promising probe of the neutrino masses \cite{Hu+98, Abazajian+15, Dvorkin+19}. The Big-Bang theory predicts the existence of a cosmic neutrino background. With nonzero masses, cosmic neutrinos can change both the Universe's expansion and the evolution of its perturbations \cite{2006PhR...429..307L,2011ARNPS..61...69W}. In early Universe, they are relativistic and act as radiation. As the Universe cools down, they become non-relativistic and behave like matter. Keeping the total matter density today fixed, non-zero neutrino masses reduce the mass fraction of baryons and cold dark matter, which leads to a later epoch of matter-radiation equality when neutrinos are typically still relativistic. In addition, the large thermal velocities of neutrinos allow them to stream out of the CDM potential wells, so they do not contribute to matter clustering, and the growth of structure is suppressed on scales smaller than their free-streaming scales. As a result, massive neutrinos leave observable imprints on both the cosmic microwave background (CMB) and large-scale structure (LSS).

The current tightest constraint on $M_{\nu}$ comes from cosmology: $M_{\nu}\lesssim 0.12$ eV at 95\% confidence level, obtained mainly by combining observations of the CMB anisotropies and the Baryonic Acoustic Oscillations (BAO) \cite{2020A&A...641A...6P,2021PhRvD.103h3533A}. Since the suppression of growth of structure by massive neutrinos is strongest on small scales and at low redshifts \cite{Brinckmann_2019,2018JCAP...03..049L}, tighter constraints are expected by including more low-redshift LSS data on small scales. 
Current and upcoming galaxy surveys such as DESI \footnote{\href{http://www.desi.lbl.gov}{http://www.desi.lbl.gov}}, PFS \footnote{\href{http://pfs.ipmu.jp}{http://pfs.ipmu.jp}}, Roman Space Telescope \footnote{\href{http://wfirst.gsfc.nasa.gov}{http://wfirst.gsfc.nasa.gov}}, Euclid \footnote{\href{http://sci.esa.int/euclid}{http://sci.esa.int/euclid}} and CSST \cite{CSST,2019ApJ...883..203G}\footnote{\href{http://nao.cas.cn/csst}{http://nao.cas.cn/csst}}, will provide high-precision measurements of the 3D clustering of galaxies. Besides, simulations have made considerable progress in modeling nonlinear structure formation \cite{Brandbyge+08,Banerjee&Dalal16,Villaescusa_Navarro_2013,Villaescusa_Navarro_2018,Castorina_2015,PhysRevD.90.045022,Castorina_2014,10.1111/j.1365-2966.2011.19488.x,2016PhRvD..93f3515U} in cosmologies with massive neutrinos. With the development in both observations and simulations, it is promising to unlock the information on small (nonlinear) scales and tightly constrain $M_{\nu}$.

However, the most commonly used statistics of LSS, the two-point correlation function or its Fourier transform, the power spectrum, is no longer sufficient to capture all statistical information on small scales (non-linear and non-Gaussian) where the effects of massive neutrinos are most prominent. Thus, statistical tools that are capable of extracting non-Gaussian information are needed to uncover the important missed information and obtain tighter constraints on $M_{\nu}$. Multiple efforts have been made in this direction. The bispectrum is demonstrated to help break the degeneracy between $M_{\nu}$ and $\sigma_{8}$ \cite{2020JCAP...03..040H} and tighten the constraints on $M_{\nu}$ \cite{Chudaykin_2019}. Void statistics have been found to be capable of capturing some characteristic effects of massive neutrinos on LSS \cite{Kreisch_2019,Massara_2015}. In order to extract the information embedded in low density regions like cosmic voids, marked power spectrum that emphasizes low density regions has been proposed to place tight constraints on $M_{\nu}$ \cite{2021PhRvL.126a1301M}. The one-point probability distribution function (PDF) of matter density field has been found to be highly complementary to the matter power spectrum on mildly non-linear scales, and help tighten the constraints on $M_{\nu}$ \cite{2020MNRAS.495.4006U}. 

In this work, we explore in a quantitive way the potential of using LSS's morphological properties to constrain $M_{\nu}$. According to Hadwidger's theorem \cite{Hadwiger_1957}, for a pattern in n-dimensional field, its morphological properties can be fully described by (n+1) Minkowski functionals (MFs). In 3D, the 4 MFs are respectively the pattern's volume, surface area, integrated mean curvature, and Euler characteristic (or genus). The MFs can principlly probe all orders of statistics \cite{1994A&A...288..697M,10.1046/j.1365-8711.1999.02912.x}. Since their introduction into cosmology by \cite{1994A&A...288..697M} in the 1990s, they have been applied to examine the Gaussianity of primordial perturbations \citep[see e.g.,][]{10.1111/j.1365-2966.2012.21103.x,2013MNRAS.435..531C,2006ApJ...653...11H,2013PhRvD..88d1302F}, to test theories of gravity \citep[see e.g.,][]{2017PhRvL.118r1301F,Shirasaki+17,Jiang,Zhong}, to probe the reionization epoch \citep[e.g.][]{Gleser+06,Chen+19}, and to investigate neutrino mass in addition to several other interesting topics \cite{2001ApJ...551L...5S,2001A&A...379..412B,2012PhRvD..85j3513K,2005ApJ...633....1P}.

In particular, previous work on probing neutrino mass with the MFs include \cite{2019JCAP...06..019M} and \cite{PhysRevD.101.063515}, which studied the MFs of the 2D weak lensing field and the 3D density field, respectively. Different from \cite{2019JCAP...06..019M} which rigorously quantified the constraints on $M_{\nu}$ from the 2D MFs, \cite{PhysRevD.101.063515} focused more on illustrating the effects of nonzero neutrino mass on the MFs of the 3D density field, and provided only indicative results for the 3D MFs' constraining power on $M_{\nu}$ by comparing a pair of simulations with neutrino mass to be zero or not \cite{Emberson_2017}. In this work, we aim at quantifying the constraints on $M_{\nu}$ from the MFs of the 3D density field, specifically, jointly with other cosmological parameters, and compare them with those from the power spectrum. We also present a detailed analysis of the effects of nonzero neutrino mass on LSS's morphology with our more realistic choice for the smoothing scales, compared to \cite{PhysRevD.101.063515} which adopted idealistic choices (specifically, $\sim$0.1 and $\sim$0.2$h^{-1}$Mpc). For this purpose, we propose a new indicator based on the MFs, i.e., the surface-area weighted average of the mean curvature, which is sensitive to the size and shape changes of voids and halos, and has been used in our analysis for the effects of peculiar velocities on LSS in \cite{2021arXiv210803851J}.

This paper is organized as follows. In Section \ref{models} we describe the two N-body simulation suites used in this work, the HADES simulation suite and the Quijote simulation suite. We then describe the measurement of Minkowski functionals in Section~\ref{sec:mfs}. The effects of neutrino mass on the MFs are detailedly analyzed in Section~\ref{effects}. The Fisher information matrix formalism used to calculate parameter constraints and the constraints on cosmological parameters from the MFs, the power spectrum, and their combination are given in Section \ref{sec:constraints}. Finally, we conclude in Section \ref{conclusions}. Convergence test is presented in Appendix \ref{sec:conver}.

\section{Models and simulations}
\label{models}

In this work, we use 400 simulations from the HADES\footnote{https://franciscovillaescusa.github.io/hades.html} suite \cite{Villaescusa_Navarro_2018} to study the effects of massive neutrinos on the MFs and use 12000 simulations from the Quijote\footnote{https://github.com/franciscovillaescusa/Quijote-simulations} suite \cite{2020ApJS..250....2V} to quantify the information content embedded in the MFs using the Fisher matrix formalism.  Specifications of the simulations used in this work can be found in Table \ref{tab:s}. The HADES simulation suite, as the precursor of the Quijote simulation suite, shares the following fiducial cosmological parameter values with the Quijote suite, which are in good agreement with the latest Planck constraints \cite{2020A&A...641A...6P}: $\Omega_{\mathrm{m}}=0.3175, \Omega_{\mathrm{b}}=0.049, h=0.6711, n_{s}=0.9624$
$\sigma_{8}=0.834, M_{\nu}=0.0$ eV and $w=-1$.  The Quijote simulations, a set of 44100 $N$-body simulations, are designed to quantify the information content of cosmological observables. Both the HADES and the Quijote simulation suites are run using the TreePM+SPH code GADGET-III \cite{10.1111/j.1365-2966.2005.09655.x} with a cosmological volume of $1\left(h^{-1} \mathrm{Gpc}\right)^{3}$ and $512^3$ CDM particles (plus $512^3$ neutrino particles for cosmologies with massive neutrinos), where cosmic neutrinos are modeled using the traditional particle-based method \cite{Viel_2010,Brandbyge_2008} and degenerate masses of massive neutrinos are assumed. The initial conditions of the HADES simulation suite are generated at $z = 99$ using the rescaling method \cite{10.1093/mnras/stw3340} employing the Zel'dovich approximation, which is also used to generate initial conditions for cosmologies with massive neutrinos and their fiducial counterparts with massless neutrinos in the Quijote suite at $z = 127$. For the initial conditions of all other cosmologies in the Quijote suite, second-order perturbation theory (2LPT) is used at $z = 127$. The gravitational evolution of particles is followed to $z = 0$ in both simulation suites, and in this work, we focus on the snapshots at this redshift.

\begin{center}
	\small
	\begin{table}[tbp]
	\begin{tabular}{cccccccccc}
		\hline Name & $M_{\nu}$ & $\Omega_{m}$ & $\Omega_{b}$ & $h$ & $n_{s}$ & $\sigma_{8}$ & $w$ & ICs & realizations \\
		\hline \hline \multicolumn{10}{c}{HADES suite} \\
		\hline Fiducial & 0.0 & 0.3175 & 0.049 & 0.6711 & 0.9624 & 0.833 & -1 & Zel'dovich & 100 \\
		& \underline{0.10} & 0.3175 & 0.049 & 0.6711 & 0.9624 & 0.815 & -1 & Zel'dovich & 100 \\
		& \underline{0.15} & 0.3175 & 0.049 & 0.6711 & 0.9624 & 0.806 & -1 & Zel'dovich & 100 \\
		\hline \hline \multicolumn{10}{c}{Quijote suite} \\
		\hline Fiducial & 0.0 & 0.3175 & 0.049 & 0.6711 & 0.9624 & 0.834 & -1 & 2LPT & 5000 \\
		Fiducial ZA & 0.0 & 0.3175 & 0.049 & 0.6711 & 0.9624 & 0.834 & -1 & Zel'dovich & 500 \\
		$M_{\nu}^{+}$ & \underline{0.1} & 0.3175 & 0.049 & 0.6711 & 0.9624 & 0.834 & -1 & Zel'dovich & 500 \\
		$M_{\nu}^{++}$ & \underline{0.2} & 0.3175 & 0.049 & 0.6711 & 0.9624 & 0.834 & -1 & Zel'dovich & 500 \\
		$M_{\nu}^{+++}$ & \underline{0.4} & 0.3175 & 0.049 & 0.6711 & 0.9624 & 0.834 & -1 & Zel'dovich & 500 \\
		$\Omega_{m}^{+}$ & 0.0 & \underline{0.3275} & 0.049 & 0.6711 & 0.9624 & 0.834 & -1 & 2LPT & 500 \\ 
		$\Omega_{m}^{-}$ & 0.0 & \underline{0.3075} & 0.049 & 0.6711 & 0.9624 & 0.834 & -1 & 2LPT & 500 \\ 
		$\Omega_{b}^{++}$ & 0.0 & 0.3175 & \underline{0.051} & 0.6711 & 0.9624 & 0.834 & -1 & 2LPT & 500 \\ 
		$\Omega_{b}^{--}$ & 0.0 & 0.3175 & \underline{0.047} & 0.6711 & 0.9624 & 0.834 & -1 & 2LPT & 500 \\ 
		$h^{+}$ & 0.0 & 0.3175 & 0.049 & \underline{0.6911} & 0.9624 & 0.834 & -1 & 2LPT & 500 \\ 
		$h^{-}$ & 0.0 & 0.3175 & 0.049 & \underline{0.6511} & 0.9624 & 0.834 & -1 & 2LPT & 500 \\ 
		$n_{s}^{+}$ & 0.0 & 0.3175 & 0.049 & 0.6711 & \underline{0.9824} & 0.834 & -1 & 2LPT & 500 \\ 
		$n_{s}^{-}$ & 0.0 & 0.3175 & 0.049 & 0.6711 & \underline{0.9424} & 0.834 & -1 & 2LPT & 500 \\ 
		$\sigma_{8}^{+}$ & 0.0 & 0.3175 & 0.049 & 0.6711 & 0.9624 &  \underline{0.849} & -1 & 2LPT & 500 \\ 
		$\sigma_{8}^{-}$& 0.0 & 0.3175 & 0.049 & 0.6711 & 0.9624 &  \underline{0.819} & -1 & 2LPT & 500 \\
		
	\end{tabular}
    \caption{\label{tab:s} The subsets of the HADES and Quijote simulation suites used in this work. Top: The HADES simulations with $M_{\nu} = 0.0,0.06,0.10$ and 0.15 eV used to study the effects of massive neutrinos on LSS. Bottom: 5000 fiducial simulations of the Quijote suite used for the estimate of the covariance matrices and 500 simulations each for 14 different cosmologies used for the calculations of derivatives of observables with respect to cosmological parameters.}
    \end{table}
\end{center}

For the HADES simulation suite, 100 massless neutrino simulations as well as 100 massive neutrino simulations for each value of $M_{\nu}$ ($M_{\nu} = 0.10, 0.15$ eV) are used. Between cosmologies with and without massive neutrinos, the primordial power spectrum amplitude $A_s$ and matter density today $\Omega_m$ are fixed. As a result, both $\sigma_8$ and $\Omega_{\mathrm{cb}}$ are smaller for cosmologies with massive neutrinos. For the Quijote simulation suite, we use 5000 fiducial simulations to obtain an accurate estimate of covariance matrices (although there are in total 15000 realizations at the fiducial cosmology, we have checked 5000 realizations are sufficient for a convergent estimate of covariance matrices), and 500 simulations at each cosmology to calculate the derivatives of observables. Between cosmologies with and without massive neutrinos, $\sigma_8$ and $\Omega_m$ are fixed, which is slightly different from the HADES suite.

\section{Measurement of Minkowski functionals}
\label{sec:mfs}

\begin{figure}[tbp]
	\centering 
	\includegraphics[width=0.8\textwidth]{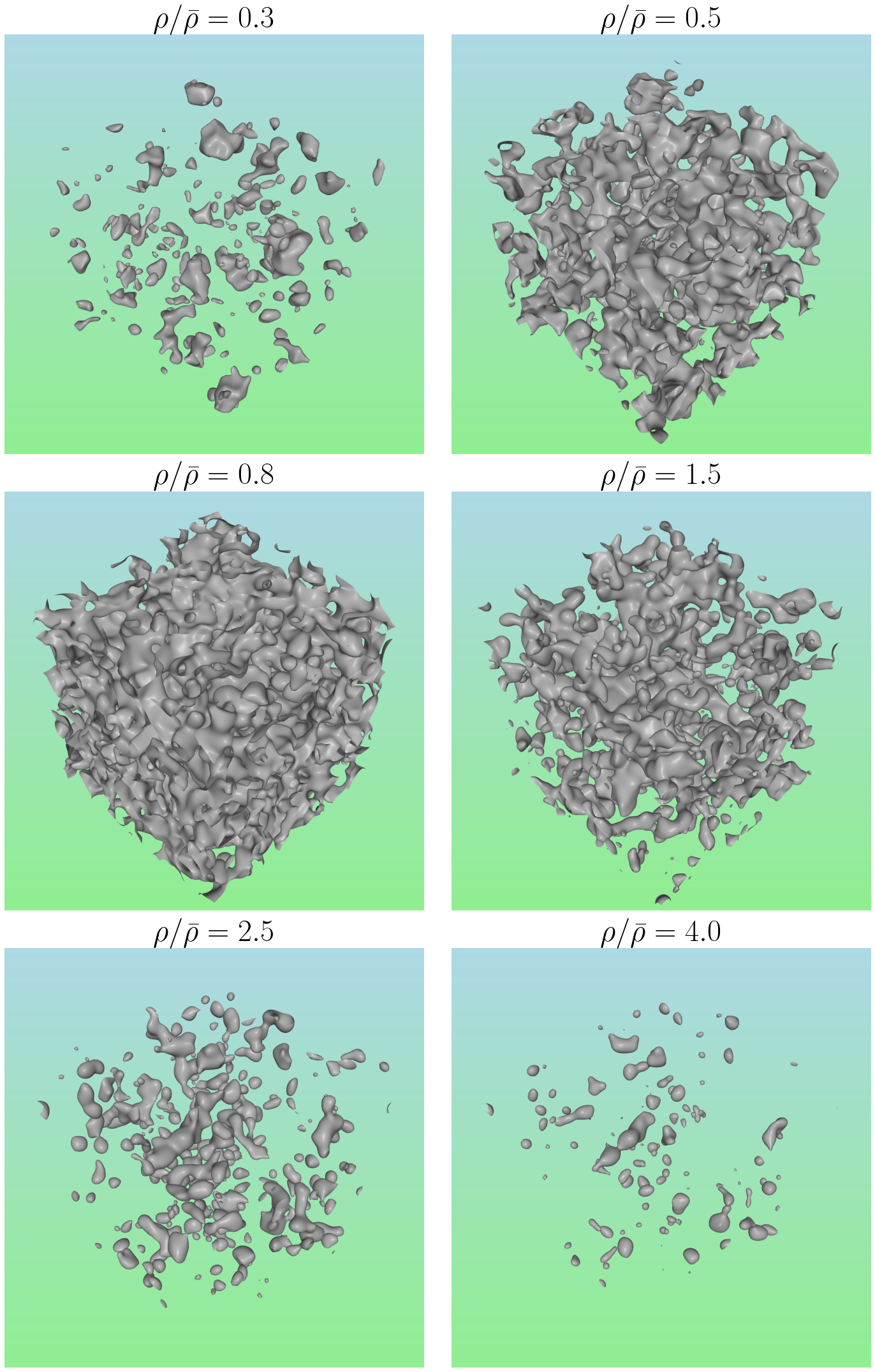}
	\caption{\label{fig:iso} Isodensity surfaces of the density field for six different density thresholds $\rho/\bar{\rho}=0.3,0.5,0.8,1.5,2.5,4.0$ in a box with volume $(250 h^{-1}\mathrm{Mpc})^3$. The density field is cropped from the original simulated box of $(1 h^{-1}\rm Gpc)^3$, and is computed from particle positions at $z=0$ for the fiducial cosmology of the HADES suite using the cloud-in-cell scheme. It is then smoothed with a Gaussian window function with scale of $R_G=5h^{-1}\mathrm{Mpc}$. The density thresholds $\rho/\bar{\rho}=0.3,0.5,0.8,1.5,2.5,4.0$ are mostly chosen to be at the extreme points of the MFs. For $\rho/\bar{\rho}=0.3$ and 0.5, isodensity surfaces mainly enclose the complement set of the excursion set, that is, regions with density below the density threshold. Therefore, their volume fraction is given by $1-V_0$.}
\end{figure}

For a spatial pattern in three-dimensional space, we can study its morphology using four Minkowski functionals, namely its volume ($V_{0}$), surface area ($V_{1}$), the integrated mean curvature ($V_{2}$) and the Euler characteristic ($V_{3}$). Instead of the functionals themselves, their spatial densities are more commonly used for convenient comparison between samples with different volumes. Therefore, in this work, we divide the four Minkowski functionals with the volume of the simulation box, thus $V_{0}$ refers to the volume fraction, while $V_{1}$, $V_{2}$ and $V_{3}$ are the surface area, the integrated mean curvature and the Euler characteristic per unit volume, respectively. To measure the MFs, the first step is to construct density fields from the positions of particles in the simulations using the cloud-in-cell mass assignment scheme on a $512^3$ grid. We have checked the effect of finite pixel size \cite{2014ApJS..212...22K} can be neglected for the density field estimated with this grid size. Since neutrinos are modeled as particles in massive neutrino cosmologies, both the total matter density field (cold dark matter + baryons + neutrinos, denoted as `m') and the cold dark matter + baryons density field (denoted as `cb') can be obtained this way.

\begin{figure}[tbp]
	\centering 
	\includegraphics[width=1.0\textwidth]{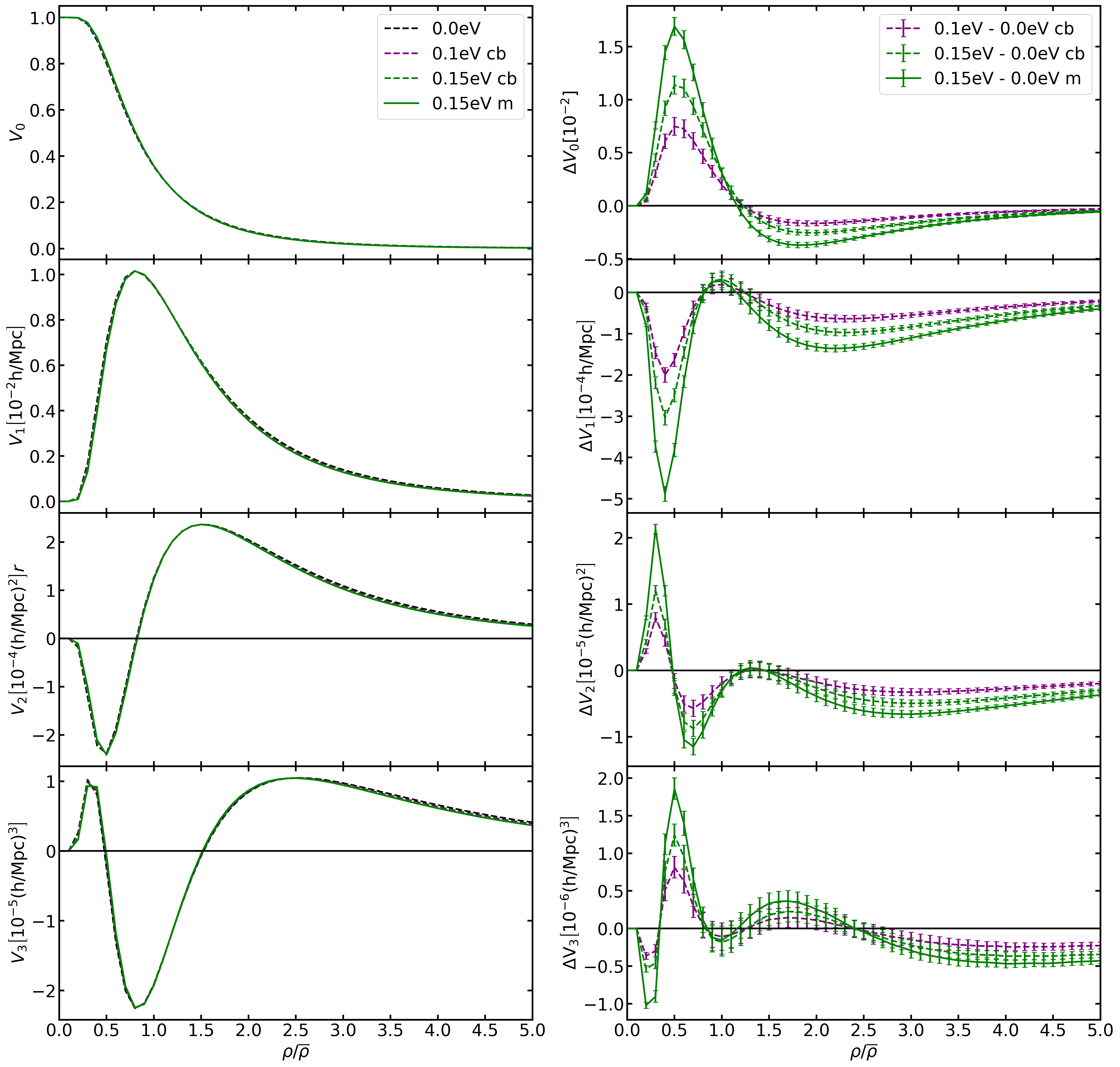}
	\caption{\label{fig:i} The MFs and differences in the MFs between different models are shown as functions of $\rho / \overline{\rho}$, which is the density threshold used for the MFs' measurements. Two density fields are considered: the `cb' (CDM + baryons, without neutrinos, dashed lines) and the `m' (CDM + baryons + neutrinos, solid lines) density field. Both the `cb' field and `m' field are smoothed with a smoothing scale of $R_G=5h^{-1}\mathrm{Mpc}$ to suppress the shot noise. We focus on $z=0$ because massive neutrinos' effects are strongest at this redshift. Left panel: The MFs for the massive and massless neutrino cosmologies, black lines for $M_{v}=0.0\,\mathrm{eV}$,  purple lines for $M_{v}=0.1\,\mathrm{eV}$ and green lines for $M_{v}=0.15\,\mathrm{eV}$. Right panel: The differences in the MFs between models with massive neutrinos and massless neutrinos. Purple lines for the difference between $M_{v}=0.1 \,\mathrm{eV}$ and $M_{v}=0.0 \,\mathrm{eV}$, and green lines for the difference between $M_{v}=0.15 \,\mathrm{eV}$ and $M_{v}=0.0 \,\mathrm{eV}$.}
\end{figure}

To suppress the shot noise, the density fields are smoothed with a Gaussian window function of radius $R_{G}$. To study the effect of massive neutrinos on the MFs using the HADES suite, a Gaussian smoothing scale of $R_G=5h^{-1}\mathrm{Mpc}$ is large enough to suppress shot noise but capable of capturing small-scale information, where the effects of massive neutrino are most significant. When quantifying the constraints on cosmological parameters from the MFs using the Quijote suite, a larger smoothing scale $R_G=10h^{-1}\mathrm{Mpc}$ is also used in addition to $R_G=5h^{-1}\mathrm{Mpc}$, to help break parameter degeneracies and tighten constraints. Then we measure the MFs for the excursion sets of the smoothed density field, which include only regions with density above a given density threshold. In Figure~\ref{fig:iso}, the surfaces of the excursion sets for six different density thresholds are presented to visualize the excursion sets. For clearer visualization, the isodensity surfaces are shown for the density field with volume $(250 h^{-1}\mathrm{Mpc})^3$ rather than the original simulated volume of $(1 h^{-1}\rm Gpc)^3$, which is cropped from the original simulation at the fiducial cosmology in the HADES suite and then smoothed with a Gaussian window function with $R_G=5h^{-1}\mathrm{Mpc}$. The density thresholds $\rho/\bar{\rho}=0.3,0.5,0.8,1.5,2.5,4.0$ are mostly chosen to be at the minimal or maximal points of the curves of MFs as shown in the left panel of Figure~\ref{fig:i}. 

 Analytical predictions for the MFs are known for Gaussian density fields \cite{Tomita_1990} and weakly non-Gaussian fields \cite{2003ApJ...584....1M,2020arXiv201104954M}. They agree with measurements from N-body simulations when the smoothing scale is large enough \cite{2004astro.ph..8428N,2020arXiv201200203M}. In this work, we focus on nonlinear scales, and the density field is thus non-Gaussian. Therefore, our findings and interpretations of the effects of massive neutrinos on the MFs are mainly based on measurements from the HADES simulations, where the primordial curvature perturbations $A_s$ and the matter density $\Omega_m$ are kept fixed between cosmologies with and without massive neutrinos. We use simulation snapshots at $z=0$, because the suppression of structure formation caused by massive neutrinos is more significant at lower redshift.  Two complementary formulae suitable for numerically calculating the MFs through differential geometry (Koenderink invariant) and integral geometry (Crofton's formula) are derived in \cite{1997ApJ...482L...1S}. In our calculations, the differences between MFs calculated with different formulae are negligible compared to those between different models. We are primarily interested in the differences between MFs of different models. Hereafter, we choose the Crofton's formula for all our calculations. The MFs for different cosmologies and the error bars of the MFs shown in figure~\ref{fig:i} are estimated with the means and the standard deviations of the MFs measured from 100 realizations, respectively.

\section{The effects of neutrino mass on the Minkowski functionals}

\label{effects}

\begin{figure}[tbp]
	\centering 
	\includegraphics[width=1.0\textwidth]{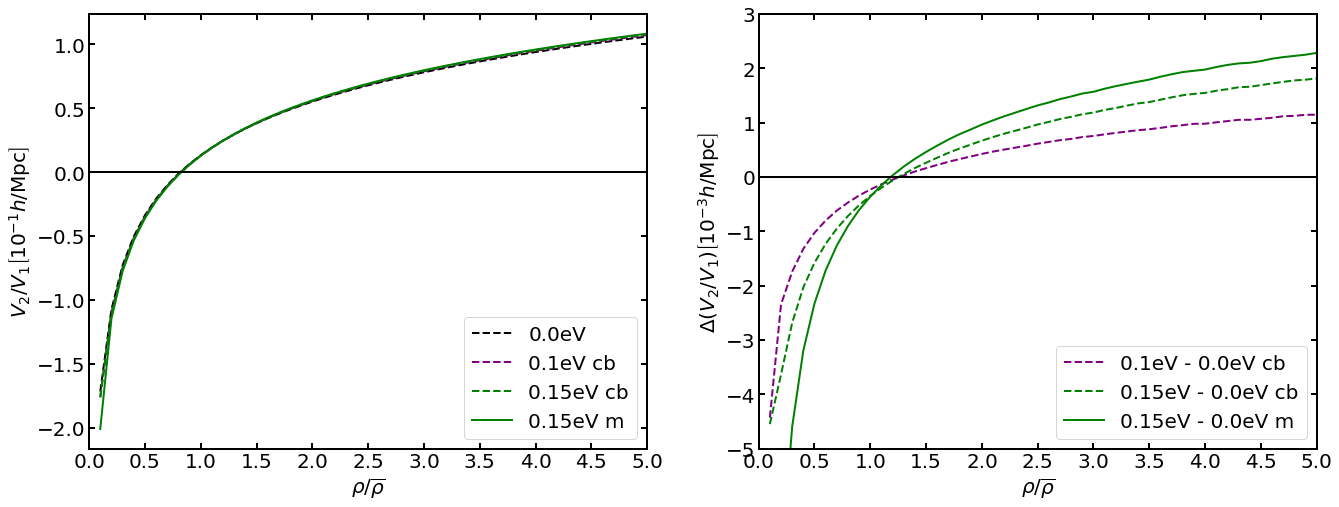}
	\caption{\label{fig:ii} The ratio of $V_{2}$ to $V_{1}$ (left panel) and the difference in $V_{2}/V_{1}$ between massive and massless neutrinos models (right panel). The line styles are the same as Figure~\ref{fig:i}. From the information presented in this figure, we can disentangle the effect of massive neutrinos on $V_1$ from their effect on the surface-area weighted average of the mean curvature, see the text for more details.}
\end{figure}

Massive neutrinos slow down structure growth on small scales; their imprints on the LSS are probed with the MFs in this section. Signatures of massive neutrinos can be seen in all of the MFs at different density thresholds. To interpret these signatures, we need to understand the curves of the MFs in the first place.  Starting by describing the symmetry of the isodensity surfaces and the MFs is helpful for understanding the MFs, which is presented in section~\ref{sec:isosurf}.  For both HADES simulations and Quijote simulations, massive neutrinos are modeled with the traditional particle-based method; the MFs can be measured for both the `m' and `cb' fields. We discuss the MFs' dependence on choice for the density field in section~\ref{sec:field}, where the MFs' dependence on the value of $M_{\nu}$ is also discussed. In section~\ref{sec:weight}, we construct a new indicator $V_2/V_1$ to help disentangle the signatures of massive neutrinos on $V_1$ and $V_2$. Like `shapefinders' constructed in \cite{1998ApJ...495L...5S,1999ApJ...526..568S}, the new indicator arises from simple, geometrical considerations and can be used to diagnose the topological properties of structures. Finally, we detailedly interpret massive neutrinos' imprints on the MFs in section~\ref{sec:nueffect}.

\subsection{The symmetry of the isodensity surfaces and the MFs}
\label{sec:isosurf}

For Gaussian random fields, the isodensity contours are statistically symmetric about $\rho/\bar{\rho}=1$ \cite{1986ApJ...306..341G,1987ApJ...321....2W},  the symmetry means the high- and low-density regions have statistically equivalent topological properties and are statistically indistinguishable from each other. The ensemble mean of the MFs of the Gaussian random field is thus symmetric at about $\rho/\bar{\rho}=1$. The non-linear gravitational evolution of the density field leads to non-Gaussianity, which produces an asymmetry in both the contours and the MFs. The point about which the isodensity contours and the MFs are symmetric is also shifted. In figure~\ref{fig:i}, we can see this point is shifted to $\rho/\bar{\rho}=0.8$.
The deviation from Gaussianity indicates a wealth of non-Gaussian information is embedded in the MFs. In the following paragraphs, we will give a detailed interpretation of the MFs, which will help understand the constraining power of the MFs.
How the contours and the MFs change as a function of $\rho/\bar{\rho}$ will be elaborated in a symmetric way, that is, the density threshold $\rho/\bar{\rho}$ goes from low- and high-density sides to $\rho/\bar{\rho}=0.8$.

On the low ($0\leq \rho/\bar{\rho} \lesssim 0.3$) and high ($ \rho/\bar{\rho} \gtrsim 2.5$) density sides,  as indicated in the panels corresponding to $\rho/\bar{\rho}=0.3$, $\rho/\bar{\rho}=2.5$ and $\rho/\bar{\rho}=4.0$ of figure~\ref{fig:iso}, contours mainly envelop void regions at low density thresholds $0 \lesssim \rho/\bar{\rho}\lesssim 0.3$ while they mostly surround high-density islands at high thresholds $2.5 \lesssim \rho/\bar{\rho}\lesssim 5$, which are composed of one or many halos (for more details about the difference between islands and halos, see \cite{2019MNRAS.485.4167P}). For voids, the density profile at the center is deepest and large voids present deeper underdensities in the core than small ones \cite{Massara_2015}.
 As the threshold is lifted from $\rho/\bar{\rho}=0$ to $\rho/\bar{\rho}= 0.3$, density contours of large voids first arise near their center and rise in size. Then contours of small voids emerge around their centers and also rise in size.
Therefore, individual voids \footnote{We are actually talking about contours of voids or halos. But for simplicity, voids, islands, and holes will refer to the contours of these structures hereafter in this context.} grow in both number and size. At the same time, adjacent voids merge into larger voids, which may have complex shapes. The situation is similar for halos: the density profile at the center is highest, and large halos have higher densities in the core than small ones \cite{1997ApJ...490..493N}. A similar process happens for islands when the threshold is lowered, density contours around massive halos first appear, and their sizes increase. Contours of less massive halos emerge a little later and then rise in size. For the density thresholds considered in this work, halos are usually included in islands, whose volume is larger than halos. When the threshold is lowered from $\rho/\bar{\rho}=5$ to $\rho/\bar{\rho}= 2.5$, isolated islands become more and larger while close islands merge into larger islands with complicated shapes.

Proceeding towards $\rho/\bar{\rho}=0.8$ from both low ($\rho/\bar{\rho}=0.3$) and high ($\rho/\bar{\rho}=2.5$)  density sides, voids and islands grow in size, they connect and merge into larger cavities and independent high-density components respectively, holes emerge in both large voids and large islands \footnote{For islands, holes emerge when adjacent ones connect and form tori with hollow cores while for voids, `holes' form when regions with relatively high density are surrounded with annular cavities.}. The number of holes keeps increasing so that it is approximately equal to the number of voids around the threshold $\rho/\bar{\rho}=0.5$ while approximately equal to the number of islands around $\rho/\bar{\rho}=1.5$. This can be inferred from the corresponding panels in figure~\ref{fig:iso} and that $V_3=0$ around both  $\rho/\bar{\rho}=0.5$ and  $\rho/\bar{\rho}=1.5$. Because $V_3$ is the sum of the isolated islands and voids minus the number of holes,  $V_3=0$ for $\rho/\bar{\rho}=0.5$ indicates voids are as abundant as holes since the number of islands can be neglected for low density threshold. On the other hand, $V_3=0$ for $\rho/\bar{\rho}=1.5$ indicates islands and holes are close in numbers since voids are rare for high density threshold. As $\rho/\bar{\rho}$ gets closer to $\rho/\bar{\rho}=0.8$, topology identities merge and form huge percolating complexes, with a large number of permeating tunnels (holes) in them. 

The symmetry between low- and high-density regions is well captured by the MFs; they are a global measure of the topological properties of structures. $V_0$ decreases and $V_1$ increases when the density threshold rises from $\rho/\bar{\rho}=0$ to $\rho/\bar{\rho}=0.3$, while both $V_0$ and $V_1$ increases when the density threshold goes from $\rho/\bar{\rho}=5$ down to $\rho/\bar{\rho}=2.5$. This is because the total volume of voids is given by $1-V_0$ while that of islands is $V_0$, therefore, when voids (islands) grow in size, $V_0$ drops (rises). For individual voids and islands, their surface area becomes larger when they grow in volume. However, for adjacent voids and islands, although they lose surface area during the merger, the increased area due to growth in volume compensates for the loss. As a result, $V_1$ increases when voids and islands become larger. When the regions enclosed by the contours are voids, the mean curvature integrated on the surface of voids is negative, because the two principal curvatures are both negative for concave voids \cite{2021MNRAS.tmp.2528L}. The integrated mean curvature of convex islands is positive while that of irregularly shaped voids and islands remains undetermined. 

On the low ($0\leq \rho/\bar{\rho} \lesssim 0.3$ and high ($ \rho/\bar{\rho} \gtrsim 2.5$) density sides,  the change in $V_2$ is dominated by that in $V_1$ (it can be dominated by the surface-area weighted average of the mean curvature $V_2/V_1$ for some threshold ranges, which will be discussed below.  For details about $V_2/V_1$, see section~\ref{sec:weight}), the integrated mean curvature of voids decreases as their surface area increases because $V_2/V_1$ is negative for voids. For halos, $V_2/V_1$ is positive, thus, $V_2$ of islands increases with $V_1$ when the threshold is lowered. Since the Euler characteristic is the sum of islands and voids minus the number of holes, we cannot specify which topological identities dominate the structure for different thresholds only from the information given by $V_3$. However, in \cite{2013JKAS...46..125P,2019MNRAS.485.4167P}, it was shown that at thresholds smaller than the first maximum point and larger than the second maximum point of $V_3$ (which is the same density threshold range considered here), the number of voids and islands are very close to the Euler characteristic, respectively. Therefore, $V_3$ rises when the number of voids or islands increases for these thresholds. 

Proceeding towards $\rho/\bar{\rho}=0.8$ from $\rho/\bar{\rho}=0.3$, $V_0$ keeps decreasing and $V_1$ keeps increasing while both $V_0$ and $V_1$ remain increasing when the density threshold continue to drop from $\rho/\bar{\rho}=2.5$ down to $\rho/\bar{\rho}=0.8$, because all topological identities become larger in volume and have larger surface area. Two extremes are presented in $V_2$ for the threshold range considered here, each of which signs the transition of the dominant topological identities in the structure, thus also signing the transition of the dominant factor that shapes the $V_2$ curve.
For $0.3\lesssim \rho/\bar{\rho}\lesssim 0.8$, $V_2$ is first dominated by $V_1$ and then by $V_2/V_1$ as $\rho/\bar{\rho}$ increases, thus $V_2$ first drops as $V_1$ increases ($V_2/V_1$ is negative), bottoms out around $\rho/\bar{\rho}=0.5$ and then rise with $V_2/V_1$. From $\rho/\bar{\rho}=2.5$ to  $\rho/\bar{\rho}=0.8$, the story is very similar, $V_2$ is also first dominated by $V_1$ and then by $V_2/V_1$ as $\rho/\bar{\rho}$ decreases, thus $V_2$ first rises as $V_1$ increases ($V_2/V_1$ is positive), peaks around $\rho/\bar{\rho}=1.5$ and then drops with $V_2/V_1$. The trend of $V_3$ simply drops when $\rho/\bar{\rho}$ approaches $0.8$, because voids and islands connect and merge into larger voids and islands, respectively. At the same time, more holes emerge in voids and islands with complicated shapes. Both processes make the value of $V_3$ smaller.  Around threshold $\rho/\bar{\rho}=0.5$, the structure is equally dominated ($V_3=0$) in numbers by voids and holes in irregularly shaped voids while around $\rho/\bar{\rho}=1.5$, by islands and holes in irregularly shaped islands. For $\rho/\bar{\rho}=0.8$, the structure is dominated by large, complicated-shaped complexes, with so many permeating holes in them that the number of holes is close to the absolute value of the Euler characteristic. However, there are still a small number of voids and islands \cite{2013JKAS...46..125P,2019MNRAS.485.4167P}.

\subsection{Dependence on choices for the density field and $M_{\nu}$}
\label{sec:field}
As mentioned in section \ref{sec:mfs}, massive neutrinos are modeled as particles both in the HADES and Quijote simulations, thus we can measure the MFs of the total matter density field (CDM + baryons + neutrinos, denoted by `m') and the CDM + baryons density field (denoted by `cb'). The density thresholds of the two density fields can be related with 
\begin{equation}
\label{thres}
\frac{\rho_{m}}{\bar{\rho}_{m}}=\frac{\rho_{\mathrm{cb}}+\rho_{\nu}}{\bar{\rho}_{\mathrm{cb}}+\bar{\rho}_{\nu}}=\frac{\rho_{\mathrm{cb}}}{\bar{\rho}_{\mathrm{cb}}} \frac{\Omega_{\mathrm{cb}}}{\Omega_{m}}+\frac{\rho_{\nu}}{\bar{\rho}_{\nu}} \frac{\Omega_{\nu}}{\Omega_{m}},
\end{equation} 
where neutrinos are denoted by `$\nu$'.  Therefore, we anticipate the effect of massive neutrinos is stronger on the `m' field than on the `cb' field. Because massive neutrino particles experience less clustering than CDM particles in cosmologies with massive neutrinos. The clustering of CDM particles is suppressed because massive neutrinos have large thermal velocities; they can stream out of CDM potential wells easily so that the total matter potential wells are shallower than those in the cosmology with $M_{\nu}=0$.  Due to their large thermal velocities, massive neutrino particles are intrinsically less clustered on scales below their free-streaming scale. Their clustering is further suppressed by the shallower total matter potential wells. Therefore, when neutrino particles are included in the density field, the effect of suppression is expected to be stronger on this `m' field than on the `cb' field \cite{Massara_2015,2020JCAP...06..032B}, massive neutrinos' effects on the MFs are thus stronger. We expect $\Delta V_i$ of the `m' density field has a larger amplitude than that of the `cb' field, which agrees with our results shown in the right panel of figure~\ref{fig:i}. 

 The sum of neutrino masses $M_{v}$ affects the amount of suppression for structure growth. As $M_{v}$ increases, the suppression is larger when $\Omega_m$ is fixed \cite{gerbino2018status}. From figure~\ref{fig:i}, we can see the difference in the MFs between massive neutrinos and massless neutrinos is approximately proportional to the sum of neutrino mass and the curves of $\Delta V_{i}$ only differ in amplitudes. As explained above, $\Delta V_{i}$ depends on whether neutrinos are considered in the density field or not and on the sum of neutrino masses. However, the overall trends of $\Delta V_{i}$ curves are the same. We thus analyze the effects of massive neutrinos on the MFs based on $\Delta V_{i}$ measured for the matter density field between the cosmology with $M_{\nu}=0.15 eV$ and the cosmology with $M_{\nu}=0.0 eV$.

\subsection{The surface-area weighted average of the mean curvature}
\label{sec:weight}
The surface integral of the mean curvature $V_2$ as a function of the density threshold $\rho / \overline{\rho}$ is given by
\begin{equation}
\label{eq:v2}
V_{2}(\rho / \overline{\rho})=\frac{1}{\mathcal{V}} \int d^{2} A(\boldsymbol{x}) v_{2}^{( \text {loc) } }(\rho / \overline{\rho}, \boldsymbol{x}),
\end{equation}
where $\mathcal{V}$ is the volume of the box and $ v_{2}^{( \text {loc) } }(\rho / \overline{\rho}, \boldsymbol{x})$ is the local mean curvature at the point $\boldsymbol{x}$. This formula shows the effect of massive neutrinos on $V_{2}$ is determined by both their effects on the surface area and on the mean curvature. However, we can disentangle the two by rewritting equation~(\ref{eq:v2}) as
\begin{equation}
V_{2}(\rho / \overline{\rho})=V_{1}(\rho / \overline{\rho})\left\langle v_{2}(\rho / \overline{\rho})\right\rangle,
\end{equation}
therefore, the ratio of $V_{2}$ to $V_{1}$ gives the surface-area weighted average of the mean curvature $\left\langle v_{2}(\rho / \overline{\rho})\right\rangle$.
The sign of $V_{2} / V_{1}$ is determined by that of $V_2$, since $V_1$ is by definition positive. The sign of $V_2$ is determined by the integrated mean curvature of dominant components in the structure. For structure dominated by voids, $V_2$ is negative, because the two principal curvatures are both negative for concave voids \cite{2021MNRAS.tmp.2528L}. The integrated mean curvature of convex islands is positive while that of irregularly shaped voids and islands with holes in them is not determined.

Since $V_{2} / V_{1}$ is a global measure of the mean curvature of the surfaces of excursion sets, it can be influenced by various changes in the structure. For example, the increase of small voids in number makes $V_{2} / V_{1}$ smaller while more small islands increase it; when voids and islands grow in size, the absolute value of $V_{2} / V_{1}$ drops; for a spheroid with two axes of length $r$ and $\lambda r$, $\vert V_{2} / V_{1} \vert$ is larger when $\vert\lambda-1\vert$ is larger~\cite{2021arXiv210803851J} for fixed $r$. $V_{2} / V_{1}$ and the difference in $V_{2} / V_{1}$ between cosmologies with massive and massless neutrinos are plotted in figure~\ref{fig:ii}, we postpone detailed interpretation of these curves to section~\ref{sec:nueffect}. We note that there is a small sharp turning point in the blue dashed line in the right panel of figure~\ref{fig:ii}, its origin is numerical rather than physical. As $\rho / \bar{\rho}\rightarrow0$, both $V_1$ and $V_2$ approach zero, thus $V_{2} / V_{1}$ suffers from numerical instability leading to the sharp turning point.

\subsection{Differences in MFs caused by massive neutrinos}
\label{sec:nueffect}

\begin{figure}[tbp]
	\centering 
	\includegraphics[width=0.85\textwidth]{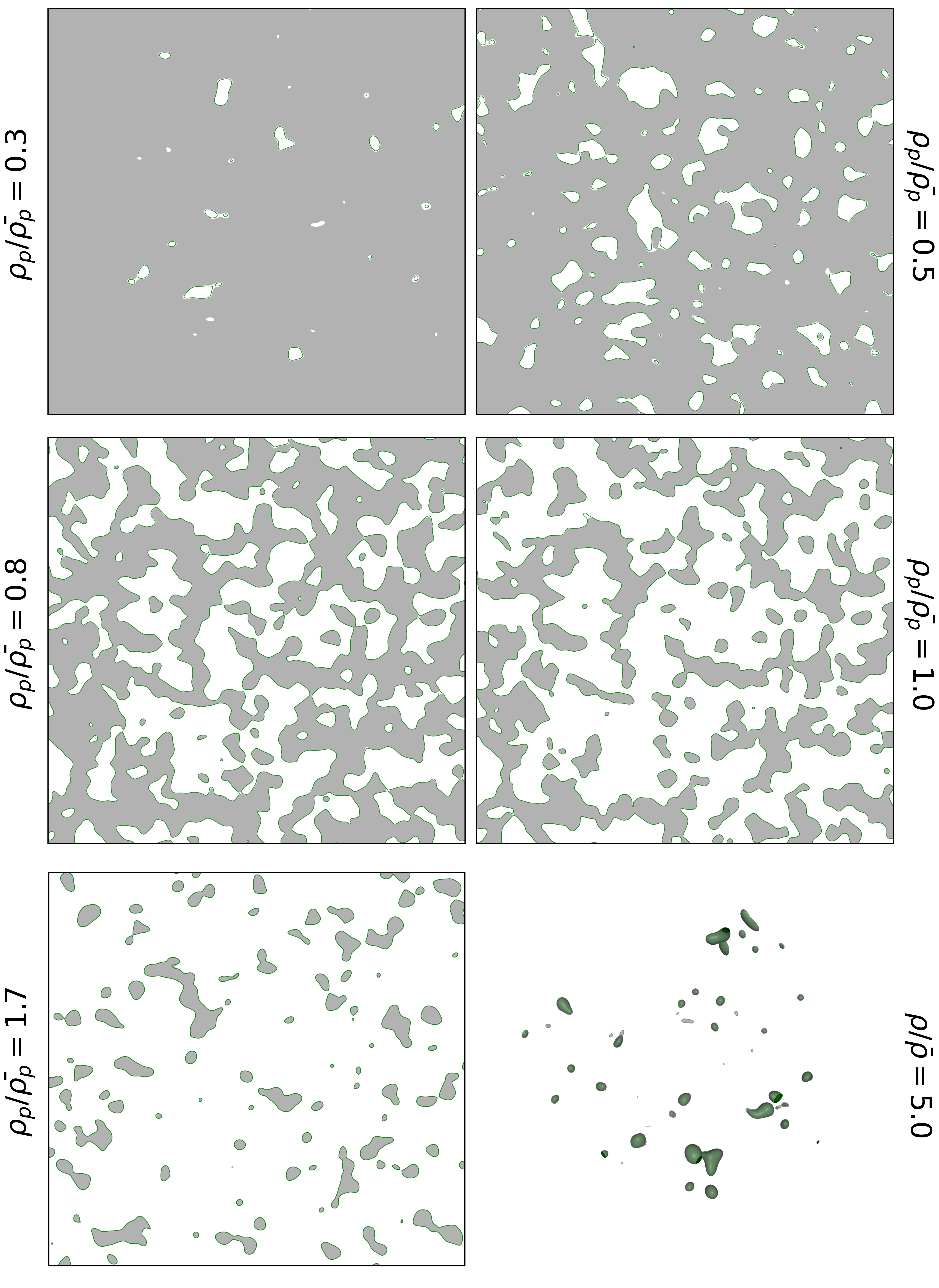}
	\caption{\label{fig:iso2d} 2D isodensity contours and excursion sets of the projected density field from a box of size $500h^{-1}$Mpc and slice thinkness $5h^{-1}$Mpc for five different density thresholds $\rho_p/\bar{\rho_p}=$ 0.3 (top-left), 0.5 (top-right), 0.8 (middle-left), 1.0 (middle-right), 1.7 (bottom-left). Grey areas depict excursion sets for the fiducial cosmology with massless neutrinos, while green lines show isodensity contours for the cosmology with $M_{\nu}=0.15 \rm eV$. In the bottom right panel, we show the 3D isodensity surfaces for $\rho/\bar{\rho}=5.0$ for a box of size $250 h^{-1}\mathrm{Mpc}$, with transparent grey contours for the fiducial cosmology and green contours for the cosmology with $M_{\nu}=0.15 \rm eV$. The density field is constructed from CDM particle positions for the fiducial cosmology and from positions of both CDM particles and massive neutrino particles for the cosmology with $M_{\nu}=0.15 \rm eV$ from the HADES suite at $z=0$ using the cloud-in-cell scheme. The field is then smoothed with a Gaussian window function with size $R_G=5h^{-1}\mathrm{Mpc}$.}
\end{figure}

Massive neutrinos suppress the growth of structure on scales smaller than their free-streaming scales, leaving imprints on corresponding excursion sets. For different ranges of the density threshold, different structures dominate (see section~\ref{sec:isosurf} for details), and massive neutrinos affect their morphological properties. In figure~\ref{fig:iso2d}, for visualization of massive neutrinos' effects, the 2D excursion set of the projected field for the fiducial cosmology is shown in grey areas while the 2D isodensity contour of the projected matter fields for the cosmology with $M_{\nu}=0.15 \rm eV$ is drawn in green lines. The five density thresholds $\rho_p/\bar{\rho_p}=$ 0.3 (top-left), 0.5 (top-right), 0.8 (middle-left), 1.0 (middle-right), 1.7 (bottom-left) are mainly chosen to visualize neutrinos' effects around extreme points of $\Delta V_3$, where $\rho_p$ and $\bar{\rho_p}$ denote the projected density field and its mean in two dimensions, respectively. These figures help understand the complicated shape of $\Delta V_i$s. For $\rho/\bar{\rho}=5.0$ (bottom-right), the 3D isodensity surface is shown for both the fiducial cosmology (transparent grey contour) and the cosmology with $M_{\nu}=0.15 \rm eV$ (green contour). Although both the excursion set and the isodensity contour defined in 2D space are only slices of those defined in 3D space, we find they give us the same insights into massive neutrinos' effect on the LSS as those given by the full 3D excursion set and the corresponding isodensity contour. On the other hand, for the intermediated thresholds, where the structure is dominated by permeating complexes with irregular shapes and a large number of holes, massive neutrinos' effect on the LSS is better visualized in 2D space than in 3D space. Therefore, in the following paragraphs, we will interpret massive neutrinos' effects on the MFs with insights given by both 2D figures and a 3D figure in figure~\ref{fig:iso2d}.

First, for $0\leq \rho/\bar{\rho} \lesssim 0.3$, the topology is dominated by voids. As is shown in the top-left panel ($\rho_p/\bar{\rho_p}=0.3$) in figure~\ref{fig:iso2d}, the number of contours of voids for massive neutrino cosmologies is smaller than the fiducial cosmology. This is understandable considering that contours for such a small density threshold mainly belong to large voids because large voids have deeper underdensities in the core than small ones; in addition, there is a smaller number of large voids in massive neutrino cosmologies \cite{kreisch2019massive,Massara_2015}. Since $V_3$ is approximately equal to the number of voids for $0\leq \rho/\bar{\rho} \lesssim 0.3$, fewer large voids in massive neutrino cosmologies mean $\Delta V_3<0$. We also notice contours of voids in massive neutrino cosmologies are smaller in size than in the fiducial cosmology, which is consistent with the density profile of voids shown in \cite{Massara_2015}: with a given height, the radius of the profile is smaller in massive neutrino cosmologies. $V_2/V_1$ is a global measure of the mean curvature of the contours; the overall smaller concave contours of voids in cosmologies with massive neutrinos lead to $\Delta (V_2/V_1) < 0$. For the threshold range considered here, we find change in $V_2$ is mainly driven by that in $V_1$. $V_2/V_1$ is negative for voids; hence a smaller surface area ($\Delta V_1<0$) leads to larger integrated mean curvature, i.e., $\Delta V_2>0$. The total volume fraction of the void regions is given by $1-V_0$, $\Delta V_0>0$ is thus consistent with our anticipation.

For $ \rho/\bar{\rho} \gtrsim 2.5$, the topology is dominated by islands. It can be seen from the bottom-right panel ($\rho/\bar{\rho}=5.0$) in figure~\ref{fig:iso2d} that the contours of islands in the massive neutrino cosmology are fewer and have an overall smaller size than in the fiducial cosmology. It is thus straightforward to infer both the total volume fraction and the total surface area of islands are reduced by massive neutrinos, that is, $\Delta V_0<0$ and $\Delta V_1<0$. For $V_2$, change in $V_2$ is mainly driven by that in $V_1$ for $ \rho/\bar{\rho} \gtrsim 2.5$. Because $V_2/V_1$ is positive for convex islands, $\Delta V_1<0$ means $\Delta V_2<0$.  We can also deduce that $\Delta (V_2/V_1) > 0$ because of the overall smaller size of islands in massive neutrino cosmologies and that small convex islands have a larger $V_2/V_1$ than large ones. Because $V_3$ is approximately equal to the number of islands for $ \rho/\bar{\rho} \gtrsim 2.5$, the fewer islands in the massive neutrino cosmology mean $\Delta V_3<0$. Since high-mass haloes prefer to live in high-density regions \cite{10.1093/mnras/282.2.347,10.1046/j.1365-8711.1999.02090.x}, the fewer and smaller high-density islands indicate the abundance of halos will be reduced by massive neutrinos, which is consistent with \cite{2018JCAP...03..049L,ichiki2012impact} where the authors found massive neutrinos cause a significant decrease in the abundance of massive halos compared to the model without the massive neutrinos.  
 
Second, for $0.3\lesssim \rho/\bar{\rho} \lesssim 0.5$, contours of small voids whose density profile is shallower in the core than large ones emerge. Contours of large voids grow in size, merge and form large irregularly shaped voids. Holes start to form in these large, irregularly shaped voids and gradually grow in quantity. From the top-right panel ($\rho_p/\bar{\rho_p}=0.5$) in figure~\ref{fig:iso2d}, we can see the total number of contours of voids is larger in cosmologies with massive neutrinos than in the fiducial cosmology; the merged voids in the fiducial cosmology are still blocked by the walls between them in massive neutrino cosmologies. This is consistent with $\Delta V_3>0$ and the larger total number of voids in massive neutrino cosmologies for voids traced by cold dark matter \cite{2018JCAP...03..049L}.  Contours of voids are generally smaller in size for massive neutrino cosmologies; thus, $\Delta (V_2/V_1) < 0$, $\Delta V_0>0$ (recall that the total volume fraction of the void regions is given by $1-V_0$) and $\Delta V_1<0$. For the density thresholds considered here, we also find change in $V_2$ is mainly driven by that in $V_1$. In this threshold range, $V_2/V_1$ is negative, $\Delta V_2>0$ because $\Delta V_1<0$. For $1.5\lesssim \rho/\bar{\rho} \lesssim 2.5$, contours of individual islands grow larger, close islands merge, and form large irregularly shaped islands where holes emerge. Notice the top-left corner of the bottom-left panel ($\rho_p/\bar{\rho_p}=1.7$) in figure~\ref{fig:iso2d}, massive neutrinos slow down the merge of close islands and thus increase the number of individual contours, $\Delta V_3>0$. The overall smaller size of these contours also means $\Delta V_0<0$, $\Delta V_1<0$ and $\Delta (V_2/V_1) > 0$. In the threshold range considered here, we find change in $V_2$ is mainly driven by that in $V_1$. $V_2/V_1$ is positive and $\Delta V_1<0$, thus $\Delta V_2<0$.

Third, when proceeding further toward $\rho/\bar{\rho}=0.8$, from the low-density side, i.e., $0.5\lesssim \rho/\bar{\rho} \lesssim 0.8$, the signs of $\Delta V_i$s are almost identical to those for $0.3\lesssim \rho/\bar{\rho} \lesssim 0.5$. The only difference is that $\Delta V_2<0$ in this threshold range while $\Delta V_2>0$ for $0.3\lesssim \rho/\bar{\rho} \lesssim 0.5$. The sign of $\Delta V_2$ is mainly determined by $\Delta (V_2/V_1)$ rather than by $\Delta V_1$ in this threshold range. Thus, even though the surface area $V_1$ reduces with massive neutrinos ($\Delta V_1 <0$.), the average mean curvature $V_2/V_1$ is more negative ($\Delta (V_2/V_1) < 0 $) thus $\Delta V_2 <0$. In the middle-left panel ($\rho_p/\bar{\rho_p}=0.8$) of figure~\ref{fig:iso2d}, massive neutrinos fill tunnels that link two adjacent voids (see examples at the bottom-right corner of this panel) in the fiducial cosmology; tunnels that connect the hollow loop in irregularly shaped voids (see one example at the middle-right of this panel) are also filled in massive neutrino cosmologies.  As can be seen from the examples shown in this panel, whenever a tunnel is filled, small convex contour patches in the fiducial cosmology will be replaced by small concave contour patches. The surface integral of the mean curvature over a small convex patch is positive, while that of a small concave patch is negative. Therefore, $V_2$ is reduced in massive neutrino cosmologies.

Proceeding further toward $\rho/\bar{\rho}=0.8$ from the high-density side, i.e., $0.8\lesssim \rho/\bar{\rho} \lesssim 1.5$, the situation is more complicated. For the density range $1.2\lesssim \rho/\bar{\rho} \lesssim 1.5$, the signs of $\Delta V_i$s are almost the same as those for the range of $1.5\lesssim \rho/\bar{\rho} \lesssim 2.5$. They only differ in $\Delta V_2$: $\Delta V_2<0$ for $1.5\lesssim \rho/\bar{\rho} \lesssim 2.5$ while the sign of $\Delta V_2$ is dependent on the value of $M_{\nu}$ for $1.2\lesssim \rho/\bar{\rho} \lesssim 1.5$. This is because $\Delta V_2$ is dominated by $\Delta V_1$ for $1.5\lesssim \rho/\bar{\rho} \lesssim 2.5$ while for $1.2\lesssim \rho/\bar{\rho} \lesssim 1.5$, whether $\Delta V_2$ is dominated by $\Delta V_1$ or $\Delta (V_2/V_1)$ depends on the value of $M_{\nu}$, hence whether $\Delta V_2>0$ is also dependent on the value of $M_{\nu}$. For the density range $0.8\lesssim \rho/\bar{\rho} \lesssim 1.2$, almost all high-density islands gradually merge into large percolating complexes with many tunnels (holes) in them, which are accompanied by a small number of independent islands. With massive neutrinos, one can see from the middle-right panel ($\rho_p/\bar{\rho_p}=1.0$) in figure~\ref{fig:iso2d} that isolated adjacent islands are connected (see one example at the top-left corner of this panel) and cavities are sealed (see one example at the bottom-left corner of this panel), which makes $\Delta V_3 <0$ while increasing the volume of irregularly shaped islands ($\Delta V_0 > 0$). In addition, after isolated islands are connected or cavities are sealed in massive neutrino cosmologies, we find convex contour patches are replaced by concave contour patches from the examples shown in this panel. As a result, both the integrated mean curvature and the surface-area weighted average of the mean curvature of the excursion set are smaller ($\Delta V_2 <0$ and $\Delta (V_2/V_1) < 0 $) in massive neutrino cosmologies. Examples can be seen in the middle-right panel that massive neutrinos also increase the size of irregularly shaped islands without changing the number of isolated islands, voids, and holes, which means that these complexes have a larger surface area $\Delta V_1 > 0$. 

Through detailed analysis of the figure~\ref{fig:i}, we find the effects of massive neutrinos on the LSS can be well captured by the MFs. In section~\ref{sec:constraints}, we show that the combination of the complete set of MFs can put tight constraints on the mass sum of neutrinos.

\section{Cosmological constraints from the MFs}
\label{sec:constraints}
In this section, we quantify the information content from the MFs for the cosmological parameters $\{\Omega_{\mathrm{m}}, \Omega_{\mathrm{b}}, \mathrm{h}, \mathrm{n}_{s}, \sigma_{8}\}$ and the sum of neutrino masses $M_{\nu}$ using the Fisher matrix formalism \cite{fisher1936use,tegmark1997karhunen}, and compare with that from the more widely-used power spectrum. 

\subsection{Fisher information matrix}  
\label{sec:fisher}
 
We use the Fisher information matrix to quantify the constraining power from the MFs and power spectrum on the cosmological parameters, which is defined as
\begin{equation}
	F_{\alpha \beta}=\left\langle-\frac{\partial^{2} \ln \mathcal{L}}{\partial \theta_{\alpha} \partial \theta_{\beta}}\right\rangle,
\end{equation}
where the likelihood $\mathcal{L}$ is assumed to be Gaussian. As suggested in \cite{2013A&A...551A..88C},  the parameter-independent covariance matrix for the observables $\boldsymbol{C}$ is used to avoid introduction of artificial information. The Fisher matrix can then be written as
\begin{equation}
	F_{\alpha \beta}=\frac{\partial \boldsymbol{\mu}}{\partial \theta_{\alpha}}^{\mathrm{T}} \boldsymbol{C}^{-1} \frac{\partial \boldsymbol{\mu}}{\partial \theta_{\beta}},
\end{equation}
where $\boldsymbol{\mu}$ is the theoretical mean for the data vector. It can be the power spectrum, the four MFs, the combination of power spectrum and MFs, or the combination of MFs with different smoothing scales etc. $\boldsymbol{C}^{-1}$ is the inverse of the covariance matrix. For $\{\Omega_{\mathrm{m}}, \Omega_{\mathrm{b}}, \mathrm{h}, \mathrm{n}_{s}$, and $\sigma_{8}\}$, the corresponding derivative is estimated as
\begin{equation}
\label{eq:para_deri}
	\frac{\partial \boldsymbol{\mu}}{\partial \theta_{\alpha}}=\frac{\boldsymbol{\mu}(\theta_{\alpha}^{+}) - \boldsymbol{\mu}(\theta_{\alpha}^{-})}{\theta_{\alpha}^{+} - \theta_{\alpha}^{-}},
\end{equation}
where $\boldsymbol{\mu}(\theta_{\alpha}^{+})$ and $\boldsymbol{\mu}(\theta_{\alpha}^{-})$ are estimated as the average of observables from 500 realizations for the models of $\theta_{\alpha}^{+}$ and $\theta_{\alpha}^{-}$, respectively. For $M_{\nu}$, although simulations with $M_{\nu}=0.2,\ 0.4 \rm eV$ are available and higher order approximations for the derivative can be constructed, we compute the derivative with respect to $M_{\nu}$ by 
\begin{equation}
\label{eq:nu_deri}
	\frac{\partial \boldsymbol{\mu}}{\partial M_{\nu}}=\frac{\boldsymbol{\mu}(M_{\nu}^{+}) - \boldsymbol{\mu}(\theta_{fid}^{ZA})}{0.1},
\end{equation}
where $\boldsymbol{\mu}(M_{\nu}^{+})$ and $\boldsymbol{\mu}(\theta_{fid}^{ZA})$ are estimated as the average of observables from 500 realizations for the models of $M_{\nu}=0.1 eV$ and fiducial model with Zeldovich initial conditions, respectively. We find Fisher matrix forecasts obtained with this approximation converge faster compared to higher order approximations. 

For the data vector of the MFs, a total of $4 \times N_{b}$ entries are combined; the four MFs share the same density threshold bins. For extensive coverage of cosmic structure and the convenience of calculations, we evenly divide the threshold range into $N_{b}=20$ bins, the two ends of which correspond to the volume fraction of 0.02 and 0.98, respectively. For the data vector of the power spectrum, a total of 80 wavenumber bins are used, up to $k_{\rm max}=0.5h\mathrm{Mpc}^{-1}$. The size of each bin is $2 \pi/L$, where $L=1h^{-1} \rm Gpc$ is the size of the simulation box. The choice of $k_{\rm max}$ is based on previous work by \cite{2020ApJS..250....2V,2021PhRvL.126a1301M}, where the authors found the constraints from the power spectrum tend to saturate for $k_{\rm max} \ge  0.5h\mathrm{Mpc}^{-1}$.

\begin{figure}[tbp]
	\centering 
	\includegraphics[width=1.0\textwidth]{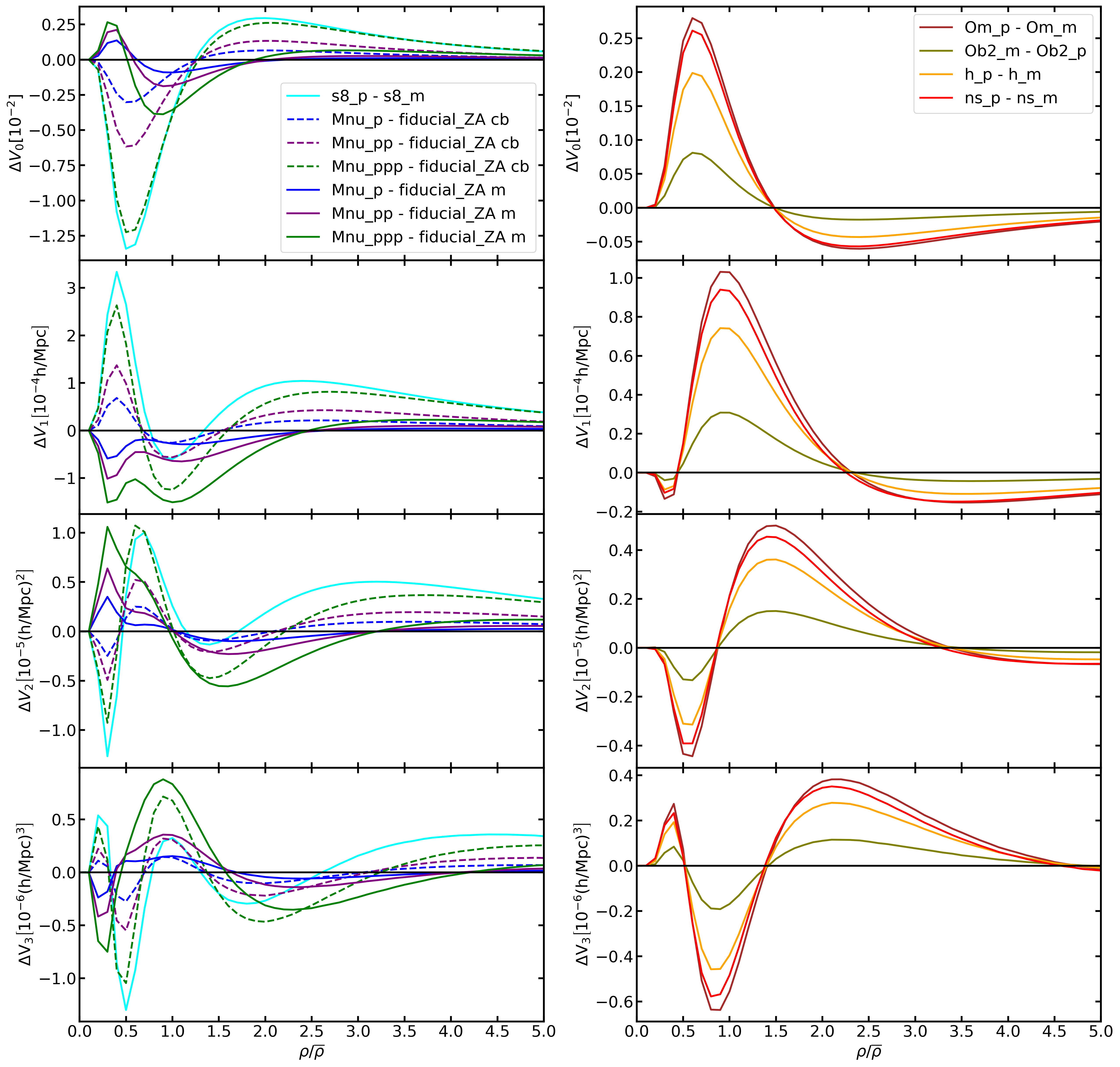}
	\caption{\label{fig:k} The differences in the MFs between different cosmologies with $R_G =  5 h^{-1}\mathrm{Mpc}$ at $z=0$, dashed lines for CDM + baryon field (cb) and solid lines for CDM + baryon + neutrinos field (m). Left panel: Blue lines for $M_{\nu}^{+}$ minus $Fiducial ZA$,  purple lines for $M_{\nu}^{++}$ minus $Fiducial ZA$, green lines for $M_{\nu}^{+++}$ minus $Fiducial ZA$, cyan lines for $\sigma_{8}^{+}$ minus $\sigma_{8}^{-}$. Right panel: Brown lines $\Omega_{m}^{+}$ minus $\Omega_{m}^{-}$, olive lines for $\Omega_{b}^{--}$ minus $\Omega_{b}^{++}$, orange lines for $h^{+}$ minus $h^{-}$ and red lines for $n_s^{+}$ minus $n_s^{-}$. See Table~\ref{tab:s} for details of these cosmological models.}
\end{figure}

In figure~\ref{fig:k}, the differences in the MFs $\Delta V_i$s between 8 pairs of cosmological models are shown, which differ from our estimates for the derivatives $\partial V_i/\partial \theta_{\alpha}$ only by a constant factor $\Delta \theta_{\alpha}$. For each pair of models, only one cosmological parameter is varied. The exact values of cosmological parameters for these models are shown in table~\ref{tab:s}.  For the cosmological parameters $\Omega_{\mathrm{m}}, \Omega_{\mathrm{b}}, \mathrm{h}, \mathrm{n}_{s}, \sigma_{8}$ and the sum of neutrino masses $M_{\nu}$, their effects on the MFs can be approximately divided into two groups (shown in the left and right panels of figure~\ref{fig:k}, respectively) according to the overall trends shared by each group. The simple cosmological dependence of the MFs has a twofold consequence. On the one hand, degeneracies exist between cosmological parameters for the MFs, limiting the constraining power of the MFs. However, one can combine the MFs at different redshifts and with different smoothing scales to break parameter degeneracies. On the other hand, one can build an emulator for the MFs based on their simple dependence on the cosmological parameters. 

From the left panel of figure~\ref{fig:k}, we note the difference in the MFs between cosmologies with massive neutrinos and with massless neutrinos from the Quijote suite is different from that from the HADES suite. This is because the amplitude of primordial curvature perturbation $A_s$ is kept fixed in the HADES suite, while $\sigma_8$ is kept fixed in the Quijote suite. For cosmologies with massive neutrinos and the fiducial cosmology with massless neutrinos, the trend of $\Delta V_i$ for the `m' field differs from that for the `cb' field. From the curves shown in the right panel of figure~\ref{fig:i} and the left panel of figure~\ref{fig:k}, one can find that the difference between $\Delta V_i$s for the `m' and `cb' density fields calculated from the Quijote suite is similar to that calculated from the Hades suite. Indeed, we have checked that they are within $1\sigma$ agreement with each other. Because the main effect of including massive neutrinos' contribution to the density field remains the same for the two different simulation settings, one can easily find that when massive neutrinos are included in the density field, the fraction of both underdense and overdense regions is lowered. In the right panel of figure~\ref{fig:k}, we find the difference in the MFs $\Delta V_i$s caused by the change of the cosmological parameters $\Omega_{\mathrm{m}}, \Omega_{\mathrm{b}}, \mathrm{h}, \mathrm{n}_{s}$ share similar trends. This indicates the presence of degeneracy between the four cosmological parameters in the MFs. Detailed analysis of their effects on the MFs is beyond the scope of the current paper, which we leave for future work.

We estimate the covariance matrices with 5000 realizations of the fiducial model. We have verified that 5000 realizations are large enough to obtain a convergent Fisher matrix and covariance matrix for the parameters. Due to uncertainties in the estimated covariance matrix $\hat{\mathbf{C}}$, the inverse of $\hat{\mathbf{C}}$ is not an unbiased estimator for $\mathbf{C}^{-1}$. Following \cite{hartlap2007your}, we remove the bias in $\hat{\mathbf{C}}^{-1}$ by 
\begin{equation}
	\label{eq:debiase}
	\mathbf{C}^{-1}=\frac{n-p-2}{n-1} \hat{\mathbf{C}}^{-1},
\end{equation}
where p is the number of observables in the data vector, and n is the number of simulations used to estimate  $\mathbf{C}$.

\begin{figure}[tbp]
	\centering 
	\includegraphics[width=1.0\textwidth]{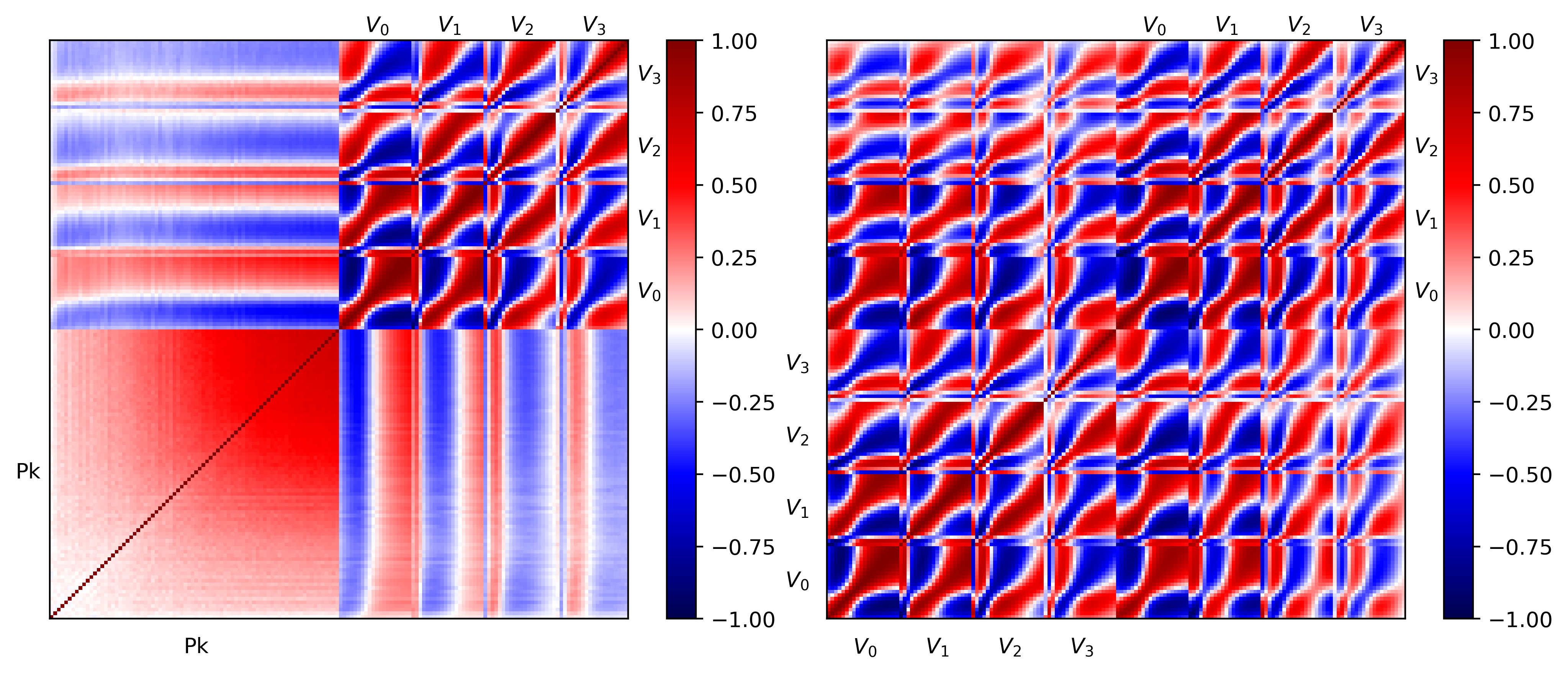}
	\caption{\label{fig:iv} Left panel: Correlation matrix of the power spectrum and the four Minkowski functionals with $R_{G}=5 h^{-1} \mathrm{Mpc}$. Right panel: Correlation matrix of the MFs with $R_{G}=5 h^{-1} \mathrm{Mpc}$ (bottom left) and the MFs with $R_{G}=10 h^{-1} \mathrm{Mpc}$ (top right). For the power spectrum, a total of 80 wavenumber bins are used, up to $k_{max}=0.5h\mathrm{Mpc}^{-1}$. The MFs share the same 20 threshold bins (thus a total of 80 MFs are used), the two ends of which correspond to volume fraction of 0.02 and 0.98, respectively. Bin values increase from left to right for both statistics. Both matrices are estimated using 5000 Quijote simulations for the fiducial model at $z=0$. The corresponding covariance matrices are used for the Fisher matrix forecasts. }
\end{figure}

In figure~\ref{fig:iv}, we show the correlation matrices of the MFs and the power spectrum. Specifically, the auto- and cross-correlations for the MFs (with $R_G=5h^{-1}\rm Mpc$) and the power spectrum are displayed on the left panel, while on the right panel, we display those for the MFs with two smoothing scales ($R_G=5h^{-1}\rm Mpc$ and a larger scale of $R_G=10h^{-1}\rm Mpc$). The correlation matrix of the power spectrum has a simple structure: cross-correlation is weak on large scales (small $k$-bins, bottom left), but gradually strengthens on smaller scales, where the correlation matrix becomes non-diagonal. The overall small correlations between the power spectrum and the MFs suggest that the MFs can provide complementary information, and more stringent constraints can be obtained with the combination of these two statistics \cite{2012PhRvD..85j3513K,2019JCAP...06..019M}. With 5000 fiducial simulations, we can accurately capture the much richer structure of the correlation matrix for the MFs. There are correlations between the MFs with different threshold bins, different orders, and different smoothing scales. The MFs at neighboring threshold bins are positively correlated because the excursion sets for neighboring bins have close morphological properties. \cite{2014MNRAS.437.2488B} proposed to use the differential MFs to obtain a more closely-diagonal covariance matrix. However, \cite{ediss27989} found although correlations between neighboring threshold bins can be alleviated with the differential MFs, the correlation matrix still exhibits a notable non-diagonal structure. Hence, in this work, we still base our Fisher matrix forecasts on the covariance matrix of the MFs themselves. As indicated in section~\ref{effects}, different orders of the MFs are anticipated to be correlated with each other. 
For example, when density thresholds are high enough, the excursion sets with higher thresholds typically have smaller total volume, smaller surface area, smaller integrated mean curvature, and smaller Euler characteristic. Since it is the same field of LSS that we smooth, we expect the MFs with $R_G=10 h^{-1}\rm Mpc$ and $R_G=5 h^{-1}\rm Mpc$ are also correlated.

For Fisher matrix forecasts, when the number of forecasted parameters is large, their covariance matrix is prone to numerical instability, and a large number of samples are needed for the estimates of both the derivatives and covariance matrix for the observables to obtain convergent parameter constraints. This phenomenon is mentioned in \cite{2021arXiv210100298B,2020JCAP...03..040H}. In Appendix~\ref{sec:conver}, we show that the number of realizations we have used to estimate the derivatives and covariance matrices is large enough to obtain convergent parameter constraints.

\subsection{Parameter constraints}
\label{sec:paraconst}
\begin{center}
	\small
	\begin{table}[tbp]
		\scalebox{0.82}[1]{
			\begin{tabular}{|c|c|c|c|c|c|c|c|c|}
				\hline Parameters & $P_{c b}$ & $M F_{c b}$ & $M F_{c b}+M F_{c b}^{\prime}$ & $P_{c b}+M F_{c b}$ & $P_{m}$ & $M F_{m}$ & $M F_{m}+M F_{m}^{\prime}$ & $P_{m}+M F_{m}$  \\
				\hline$\Omega_{m}$ & 0.0665 & 0.092 &0.0659 & 0.0207 & 0.1001 & 0.0816 & 0.0649 & 0.0139 \\
				\hline$\Omega_{b}$ & 0.018 & 0.0292 & 0.0193 & 0.0109 & 0.0397 & 0.0282 & 0.0191 & 0.01 \\
				\hline$h$ & 0.1982 & 0.3026 & 0.2117 & 0.1128 & 0.5153 & 0.3183 & 0.2215 & 0.1007 \\
				\hline$n_{s}$ & 0.1336 & 0.1078 & 0.0608 & 0.0715 & 0.4902 & 0.0982 & 0.0606 & 0.0602 \\
				\hline$\sigma_{8}$ & 0.1044 & 0.0262 & 0.0172 & 0.0204 & 0.013 & 0.0017 & 0.0016 & 0.0014 \\
				\hline$M_{\nu}$ & 1.7695 & 0.4351 & 0.2852 & 0.3381 & 0.8415 & 0.0177 & 0.0146 & 0.0133 \\
				\hline
			\end{tabular}
		}
		\caption{\label{tab:j} Marginalized errors on cosmological parameters obtained with Fisher matrix analysis. Constaints from the power spectrum ($k_{max}=0.5h\mathrm{Mpc}^{-1}$), the Minkowski functionals with $R_{G} = 5 h^{-1} \mathrm{Mpc}$, the combination of the MFs with $R_{G} = 5 h^{-1} \mathrm{Mpc}$ and $R_{G} = 10 h^{-1} \mathrm{Mpc}$, and the combination of the power spectrum and the MFs with $R_{G} = 5 h^{-1} \mathrm{Mpc}$ for the cold dark matter + baryons (denoted by subscript `$cb$') and matter (denoted by subscript `$m$') density fields are shown here.} 
	\end{table}
\end{center}

\begin{figure}[tbp]
	\centering 
	\includegraphics[width=1.0\textwidth]{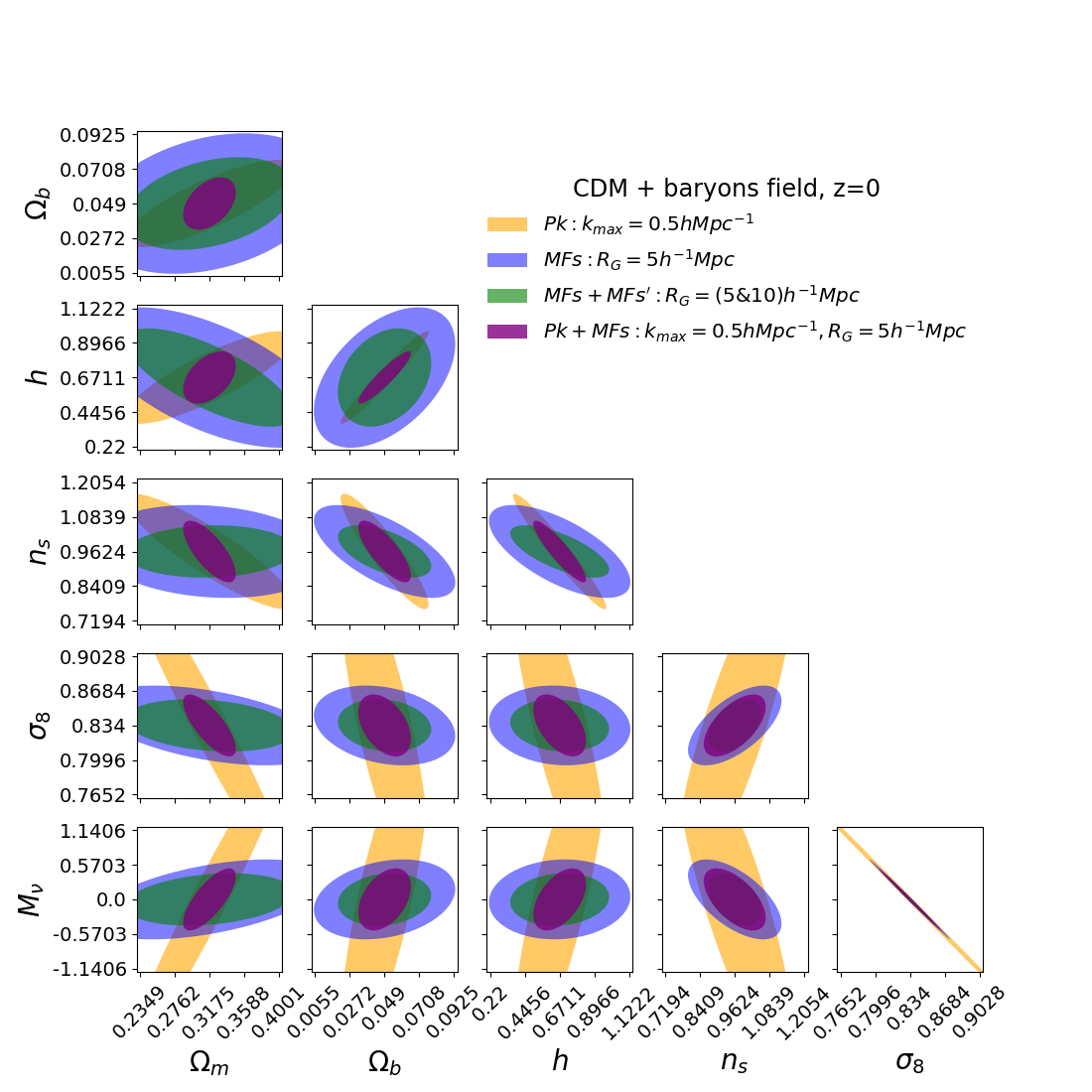}
	\caption{\label{fig:m} Constraints on cosmological parameters from observables in real space at $z=0$. Orange contours for the CDM + baryon power spectrum with $k_{max}=0.5 h \mathrm{Mpc}^{-1}$ (80 $k$ bins), blue contours for the MFs of the CDM + baryon density field with smoothing scale $R_G=5 h^{-1}\mathrm{Mpc}$ (20 density threshold bins for the four MFs, 80 observables in total), green contours for the combination of the MFs with smoothing scales $R_G=5 h^{-1}\mathrm{Mpc}$ and $R_G=10 h^{-1}\mathrm{Mpc}$, purple contours for the combination of the power spectrum and the MFs ($R_G=5 h^{-1}\mathrm{Mpc}$). }
\end{figure}

\begin{figure}[tbp]
	\centering 
	\includegraphics[width=1.0\textwidth]{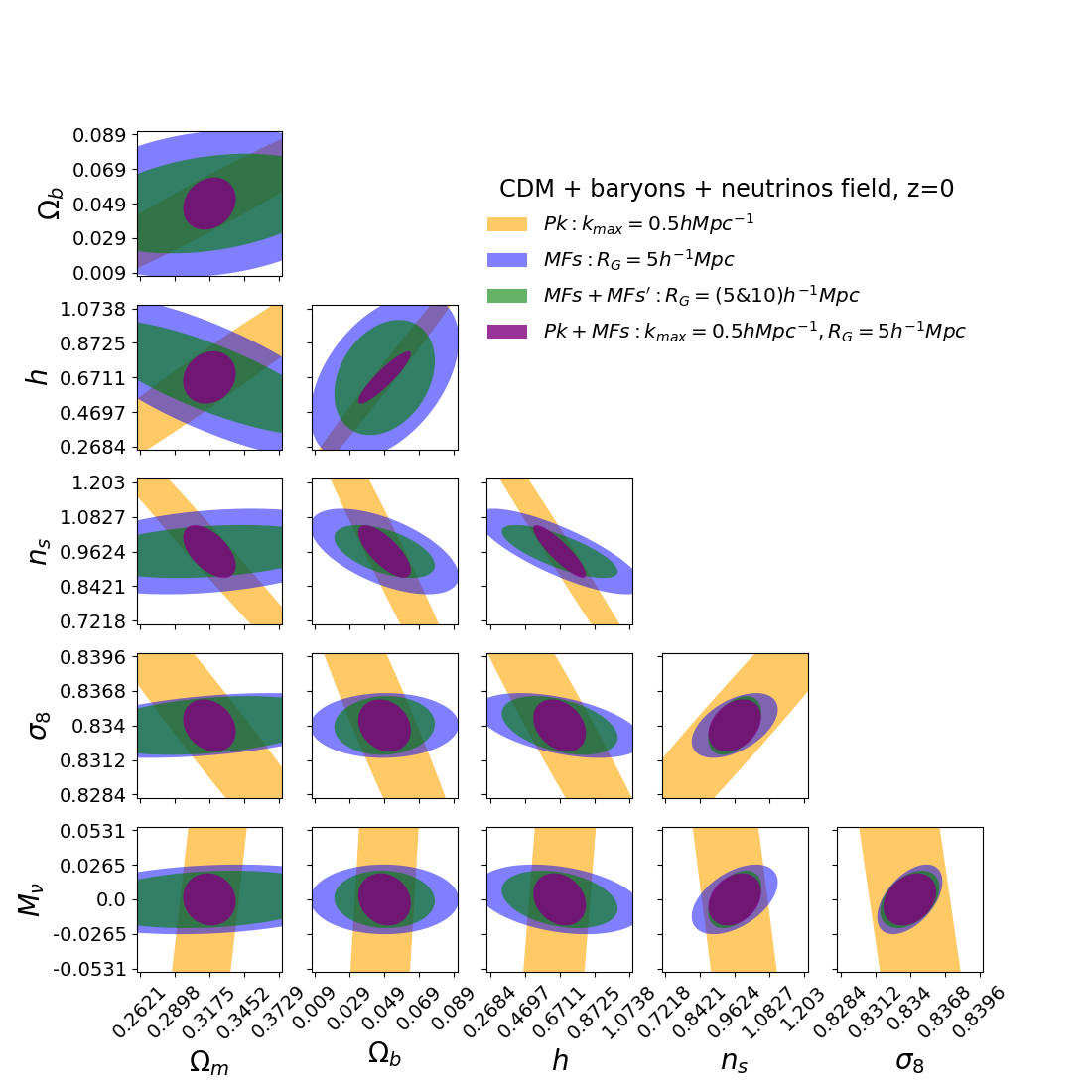}
	\caption{\label{fig:l} Constraints on cosmological parameters from observables in real space at $z=0$. Orange contours for the matter power spectrum with $k_{max}=0.5 h \mathrm{Mpc}^{-1}$ (80 $k$ bins), blue contours for the MFs of the matter density field with smoothing scale $R_G=5 h^{-1}\mathrm{Mpc}$ (20 density threshold bins for the four MFs, 80 observables in total), green contours for the combination of the MFs with smoothing scales $R_G=5 h^{-1}\mathrm{Mpc}$ and $R_G=10 h^{-1}\mathrm{Mpc}$, purple contours for the combination of the power spectrum and the MFs ($R_G=5 h^{-1}\mathrm{Mpc}$).}
\end{figure}

\begin{figure}[tbp]
	\centering 
	\includegraphics[width=1.0\textwidth]{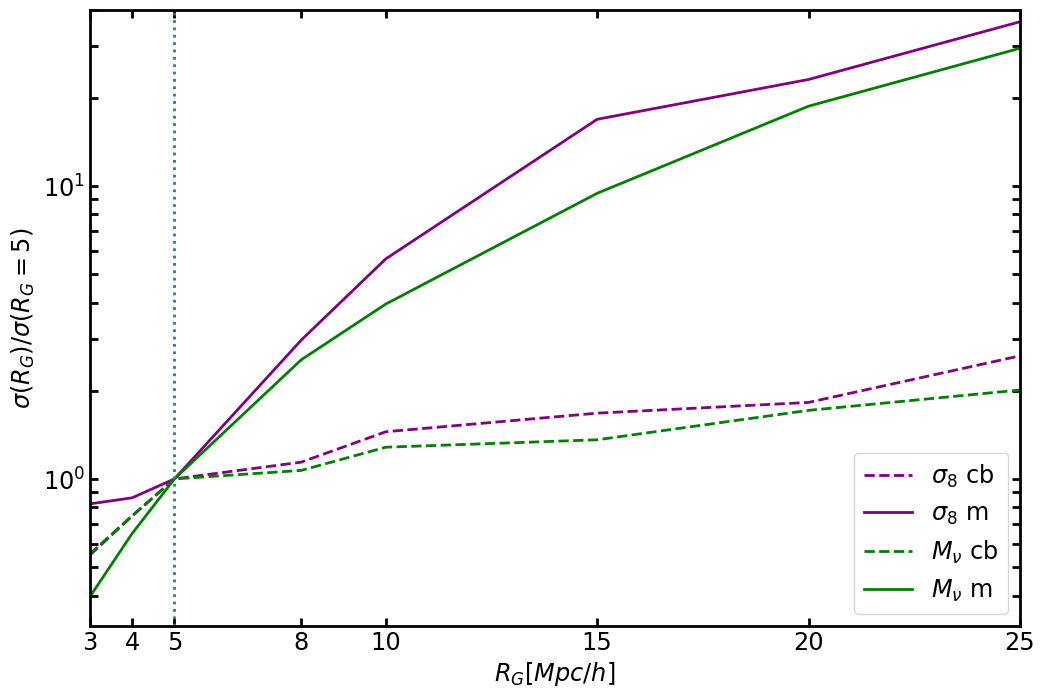}
	\caption{\label{fig:sigma_R_G} The ratios of constaints from the MFs with a smoothing scale of $R_G$ to those from the MFs with $R_G=5 h^{-1} \rm Mpc$ for $\sigma_8$ (purple lines) and $M_{\nu}$ (green lines) and for the `cb' field (dashed lines) and `m' field (green lines). Vertical blue dotted line denotes $R_G=5 h^{-1} \rm Mpc$. For both the constraint on $\sigma_8$ and on $M_{\nu}$, a stronger dependence on $R_G$ can be seen for the `m' field.}
\end{figure}

We show the marginalized $68\%$ confidence contours for different pairs of parameters obtained from our Fisher matrix calculations for the `cb' density field in figure~\ref{fig:m} and for the `m' density field in figure~\ref{fig:l}. The marginalized errors on cosmological parameters for the two different density fields are listed in Table~\ref{tab:j}. For both fields, we present the constraints from the power spectrum, the MFs with  $R_G = 5 h^{-1}\rm Mpc$, the combination of the two, and the combination of the MFs with $R_G = 5 h^{-1}\rm Mpc$ and $R_G = 10 h^{-1}\rm Mpc$. We note that the information content from the MFs does not saturate with our current choices for the threshold bins, smoothing scales, and redshifts. The constraints from the MFs can be improved with smaller smoothing scales, which is shown in figure~\ref{fig:sigma_R_G} where the deterioration of the constraint on $\sigma_8$ and on $M_{\nu}$ with the increase of $R_G$ for both the `cb' and `m' fields can be easily seen. However, due to the limitation of the numerical resolution of the Quijote simulation, we mainly focus on the smoothing scales $R_G \geq 5 h^{-1}\rm Mpc$ in this work. The area of confidence contours can be further reduced by increasing $N_b$, and/or combining the MFs with more different smoothing scales, and/or combining the MFs at different redshifts. We leave investigations on optimal choices for the threshold binning, the combination of different smoothing scales, and different redshifts to obtain the tightest constraints on $M_{\nu}$ for future work.

For the `cb' density field, the MFs provide relatively weak constraints on $\Omega_m$, $\Omega_b$ and $h$, while tighter constraint on $n_s$, and even tighter constraints on $\sigma_8$ and $M_{\nu}$, than the power spectrum. Specifically, the MFs' constraints on $M_{\nu}$, $\sigma_8$ and $n_s$ are better by a factor of 4, 4 and 1.2, respectively. The MF $V_0$ has a direct relation ($V_0$ is the complementary cumulative probability distribution function) with the one-point probability distribution function (PDF), whose constraining power on cosmological parameters was studied in detail in \cite{2020MNRAS.495.4006U}. The authors found that the PDF is weaker compared to the power spectrum in constraining $n_s$. As expected, we find that $V_0$ alone places a weak constraint on $n_s$.  We also find $V_1$, $V_2$ and $V_3$ alone all place weaker constraints on $n_s$ than $P(k)$. However, the four MFs combined together, can place a comparative or even tighter constraint on $n_s$, because the combination breaks degeneracies between $n_s$ and other parameters. For both $\sigma_8$ and $M_{\nu}$, as shown in figure~\ref{fig:k}, degeneracies between these two parameters and other parameters are relatively weak, thus resulting in much stronger constraints on both $\sigma_8$ and $M_{\nu}$. Although the MFs still exhibit the well-known degeneracy between $\sigma_8$ and $M_{\nu}$ for the 'cb' density field \cite{2021arXiv210807821V}, they can probe structure information in both the underdense and overdense regions at the same time, and principally can extract non-Gaussian information contained in the higher-order statistics of the density field. These help to break degeneracies and obtain tighter constraints \cite{2021ApJ...919...24B}. 

The degeneracy directions between cosmological parameters are different for constraints from the MFs with different smoothing scales. When the two smoothing scales $R_G = 5 h^{-1}\rm Mpc$ and $R_G = 10 h^{-1}\rm Mpc$ are combined, we find the orientation of the error ellipse slightly rotates, and its axis shrinks by a factor of about $1/3$ compared to the results for the MFs with a single smoothing scale of $R_G = 5 h^{-1}\rm Mpc$. Because the correlations between the power spectrum and the MFs are relatively small, and the two probes generally show very different degeneracies among cosmological parameters, their combined probe breaks parameter degeneracies, and the resulting constraints are much tighter compared with individual probes alone. For $\Omega_m$, $\Omega_b$ and $h$, the constraints from the combination of the power spectrum and the MFs with $R_G = 5 h^{-1}\rm Mpc$ are tighter than those from the combination of the MFs with the two different smoothing scales, due to the better sensitivity of $P(k)$ to these parameters. As expected, the combination of the MFs with the two smoothing scales is stronger in constraining $n_s$, $\sigma_8$ and $M_{\nu}$, due to the relatively stronger constraining power of the MFs on these parameters. 

For the `m' density field, as can be found from the left panel of figure~\ref{fig:k}, massive neutrinos affect the MFs in a distinct way, which considerably reduces the degeneracy between $M_{\nu}$ and $\sigma_8$. Hence the constraints on both $M_{\nu}$ and $\sigma_8$ are significantly improved, specifically by a factor of 25 and 15, respectively, compared to those from the MFs for the `cb' density field.  We note these improvements are similar to those for the constraints from the marked power spectrum\cite{2021PhRvL.126a1301M}, the wavelet scattering transform\cite{2021arXiv210807821V}, and the wavelet moments \cite{2022arXiv220407646E} when the `m' density field is used rather than the `cb' density field. For the `m' field, the extra information arises from the difference in non-linear evolution between a cosmology with 1 fluid (CDM) and 2 fluids (CDM + neutrinos), therefore much information exists beyond the power spectrum in the 3-dimensional matter field \cite{2021arXiv210804215B}.  As for the constraints from P(k), marginalized errors on $M_{\nu}$ and $\sigma_8$ improve by a factor of 2 and 8 respectively, which agrees with findings from \cite{2021PhRvL.126a1301M,2021arXiv210807821V}.  We note that the MFs give better constraints on all 6 cosmological parameters than the power spectrum when considering the `m' density field. We find the constraining power of the MFs on $M_{\nu}$ and $\sigma_8$ are more dominated by small-scale information of the matter density field, and the degeneracy between $M_{\nu}$ and $\sigma_8$ is also weaker on smaller scales. Therefore, when the information from the MFs with $R_{G} = 10 h^{-1} \rm Mpc$ is added, the improvements on the constraints of both $M_{\nu}$ and $\sigma_8$ are small, while those on other parameters are about a factor of $1/3$ (similar to previous findings for the 'cb' density field). Finally, the combination of the power spectrum and the MFs with $R_{G} = 5 h^{-1} \rm Mpc$ gives the tightest constraints on all 6 cosmological parameters among the four cases we study.

Focusing on the constraints on $M_{\nu}$, we find the MFs with a smoothing scale of $5 h^{-1} \rm Mpc$ give tighter constraints than the power spectrum. The constraints are better by a factor of $4$ when using the  `cb' density field, and a factor of $48$ when using the `m' density field. When the MFs are combined with the power spectrum, they can improve the constraints on $M_{\nu}$ from the power spectrum by a factor of  5 for the `cb' density field and 63 for the  `m' density field. Notably, when the `m' density field is used, constraint on $M_{\nu}$ from the MFs can reach $0.0177$eV, while the combination of the MFs and power spectrum can tighten this constraint to be $0.0133$eV, a $4.5\sigma$ significance on detecting the minimum sum of the neutrino masses. Note our forecasted survey volume is $1h^{-1} \rm Gpc^3$. This strong constraining power on $M_{\nu}$ from the MFs is similar to the marked power spectrum proposed by the recent work of \cite{2021PhRvL.126a1301M}.  

At the end of this section, we show the constraints on $\sigma_8$ and $M_{\nu}$ from the MFs as a function of $R_G$ for both the `cb' and `m' fields in figure~\ref{fig:sigma_R_G}. We find two reasons for the deterioration of constraints when the smoothing scale is increased. First, larger smoothing scales smear out more smaller scale structure information, thus, constraints tend to get worse. Second, the degeneracy between $\sigma_8$ and $M_{\nu}$ is generally more severe when larger $R_G$s are adopted. As can be seen, both parameter constraints have a stronger dependence on the smoothing scale for the `m' field. This is understandable: for the `m' field, we find the degeneracy between $\sigma_8$ and $M_{\nu}$ gets severe more quickly when increasing $R_G$, thus constraints on both parameters deteriorate more rapidly. 

\section{Discussions and conclusions}
\label{conclusions}
Massive neutrinos leave distinct imprints on LSS. Although it may be unclear which statistics is the most powerful to capture these signatures, and places the tightest constraints on $M_{\nu}$, we know that observables on small scales contain the key information, where these signatures are most pronounced.

In this work, we study the potential of constraining neutrino mass using the morphological properties of LSS, represented by the four MFs, which are complementary to traditional statistics such as the power spectrum and higher order statistics. We illustrate the effects of massive neutrinos on the MFs of LSS based on the Hades simulation, and proceed to quantify the information content embedded in the MFs by using the Fisher matrix formalism based on the Quijote simulation. We find the effects of massive neutrinos on LSS can be well captured by the MFs. The MFs can not only extract information consistent with statistics such as halo mass function and void size function, but also capture the distinct effects of massive neutrinos on intermediate density regions with $\rho/\bar{\rho}$ around $1$, in addition to providing extra topological information. Hence the MFs provide complementary information, and we find their combination with the power spectrum places tight constraints on $M_{\nu}$, $\sigma_8$, and other cosmological parameters. 

We have measured the MFs and power spectrum for both the `cb' field (without neutrino contribution) and the `m' field (with neutrino contribution). For the `cb' field, the MFs place stronger constraints on $n_s$, $\sigma_8$ and $M_{\nu}$ than the power spectrum, regardless of the strong degeneracy between $\sigma_8$ and $M_{\nu}$; while for the `m' field, the $\sigma_8$ and $M_{\nu}$ degeneracy is broken with the MFs, and we find they give stronger constraints on all 6 cosmological parameters than the power spectrum. Focusing on $M_{\nu}$, we find the constraint from the MFs, specifically, with a smoothing scale of $R_G=5 h^{-1}$Mpc, is $4$ times better than that from the power spectrum when using the  `cb' field, and $48$ times better when using the `m' field. When the MFs are combined with the power spectrum, they can improve the constraint on $M_{\nu}$ from the latter by a factor of  5 for the `cb' field and a factor of 63 for the  `m' field. Notably, when the `m' field is used, constraint on $M_{\nu}$ from the MFs can reach $0.0177$eV with a volume of $1(h^{-1}\rm Gpc)^3$, while the combination of the MFs and power spectrum can tighten this constraint to be $0.0133$eV, a $4.5\sigma$ significance on detecting the minimum sum of the neutrino masses. Combination with the MFs also tightens the constraints from the power spectrum on other parameters, specifically, $\left\{\Omega_{m}, \Omega_{b}, h, n_{s}, \sigma_{8}\right\}$ by a factor of 3.2, 1.7, 1.8, 1.9 and 5.1 for the `cb' field, and a factor of 7.2, 4.0, 5.1, 8.1 and 9.3 for the `m' field.

Our constraints on neutrino mass from the MFs can be further improved. Besides going to even smaller scales, using more optimal bin choices for the MFs can also improve the constraints. For example, we find using more threshold bins can lead to better constraints. Next, although combining the MFs with different smoothing scales may not be the most efficient way to extract information on structure formation from multi-scales, it helps to break parameter degeneracies, for the degeneracy directions from MFs with different smoothing scales are usually different. Thus, the combination of more smoothing scales is expected to enhance the constraints. Then, combining MFs at multiple redshifts can tighten the constraints too, as suggested by previous studies with weak lensing MFs \cite{2012PhRvD..85j3513K,2019JCAP...06..019M}. The analysis of the Quijote simulation with the one-point PDF of the matter density field \cite{2020MNRAS.495.4006U} can also give us some insights on how the constraints will change when more redshift results have been added, because the $0^{\rm th}$ order MF $V_0$ has a direct relation ($V_0$ is the complementary cumulative probability distribution function) to the one-point PDF. The combination of multiple redshifts leads to much tighter constraints due to degeneracy breaking, as is shown in Figure 15 of \cite{2020MNRAS.495.4006U}. In a preliminary study, we obtain similar findings when combining MFs at multiple redshifts. Finally, a larger survey volume can reduce the errors further. In~\cite{Jiang}, we find cosmological constraints from the MFs scale with the survey volume $V$ roughly as $1/\sqrt{V}$. Current and upcoming galaxy surveys such as DESI, Euclid, and CSST, will cover tens of $(h^{-1}\rm Gpc)^3$. Should we forecast for a survey volume like this, the constraints we obtain would be better by a factor $\sim10$. Thus, we expect these surveys have a good chance of detecting masses of neutrinos with a high significance level through measurements of the MFs.

We note our results are obtained from the distribution of the simulated CDM particles for the `cb' density field and also massive neutrino particles for the `m' density field. In reality, the `m' density field cannot be observed directly, although it gives much tighter constraints than the `cb' field. Generally speaking, weak lensing can probe the projected 2D `m' field, but efforts are still needed to reconstruct the 3D `m' field from it to allow the 3D MFs measurable. As for the `cb' field, galaxy and other surveys can trace it, but are subject to tracer biases and systematic effects such as redshift-space distortions, the Alcock-Paczynski effects, etc. Currently, in an ongoing work~\cite{Wei_HOD}, we are using the MOLINO mock galaxy catalogs \cite{2021JCAP...04..029H} constructed from the Quijote simulation to forecast constraints from the MFs of galaxy distribution with the presence of halo occupation distribution model parameters. From another ongoing work, we find the MFs for the distribution of dark matter halos are more competitive in constraining f(R) gravity than those for the dark matter density field \cite{Jiang}. A recently published work of ours finds that the MFs of matter distribution measured in redshift space place tighter constraints on cosmological parameters than the redshift-space power spectrum \cite{2021arXiv210803851J}. Both these preliminary findings are encouraging. Besides the above-mentioned systematics, when applying the MFs to real surveys, one has to take care of the irregular shapes of the masked regions and survey boundaries. However, the effects of these systematics plus varying radial and angular selection functions can be corrected, and the MFs for real galaxy catalogs can be unbiasedly reconstructed \cite{2014MNRAS.437.2488B,2021arXiv211006109A}.  In future work, we plan to investigate all these systematic effects comprehensively, and extract the constraints on neutrino mass by measuring the MFs from real surveys.  

Finally, another possible source of systematic is the baryonic effects, which are important on small scales. It is unclear how baryonic effects change the MFs of LSS and the derived constraints on neutrino mass. However, \cite{Villaescusa_Navarro_2018,2017MNRAS.471..227M} found the effects of baryonic physics and free-streaming of massive neutrinos on LSS can be treated independently. In an ongoing work \cite{Zhong}, we are studying how baryonic effects influence the MFs of the spatial distribution of neutral hydrogen and their constraints on f(R) gravity. We plan to extend this study to the MFs of 3D LSS and their constraints on neutrino mass in future work.

The MFs share common systematic effects as other statistics of LSS, though on linear scales, they are supposed to be more robust to some of the systematic effects such as nonlinear gravitational evolution and tracer bias \cite{2003ApJ...584....1M,2014MNRAS.437.2488B,2017PhRvL.118r1301F}, compared to the widely-used 2-point statistics. However, different observables generally have different sensitivities to different systematics. By comparing constraints from different observables, we can check the robustness of the results, and combining them can lead to more stringent constraints. Therefore, the MFs of LSS can definitely help achieve a tight and robust constraint on neutrino mass from cosmology.

\acknowledgments
We thank Francisco Villaescusa-Navarro for helpful discussions and Yu Yu, Pengjie Zhang, and Jia Liu for useful conversations. This work is supported by the National Natural Science Foundation of China Grants No. 12173036, 11773024, 11653002, 11421303, by the National Key R\&D Program of China Grant No. 2021YFC2203100, by the China Manned Space Project Grant No. CMS-CSST-2021-B01, by the Fundamental Research Funds for Central Universities Grants No. WK3440000004 and WK3440000005, and by the CAS Interdisciplinary Innovation Team.

\appendix
\section{Convergence test}
\label{sec:conver}

\begin{figure}[tbp]
	\centering
	\includegraphics[width=1.0\textwidth]{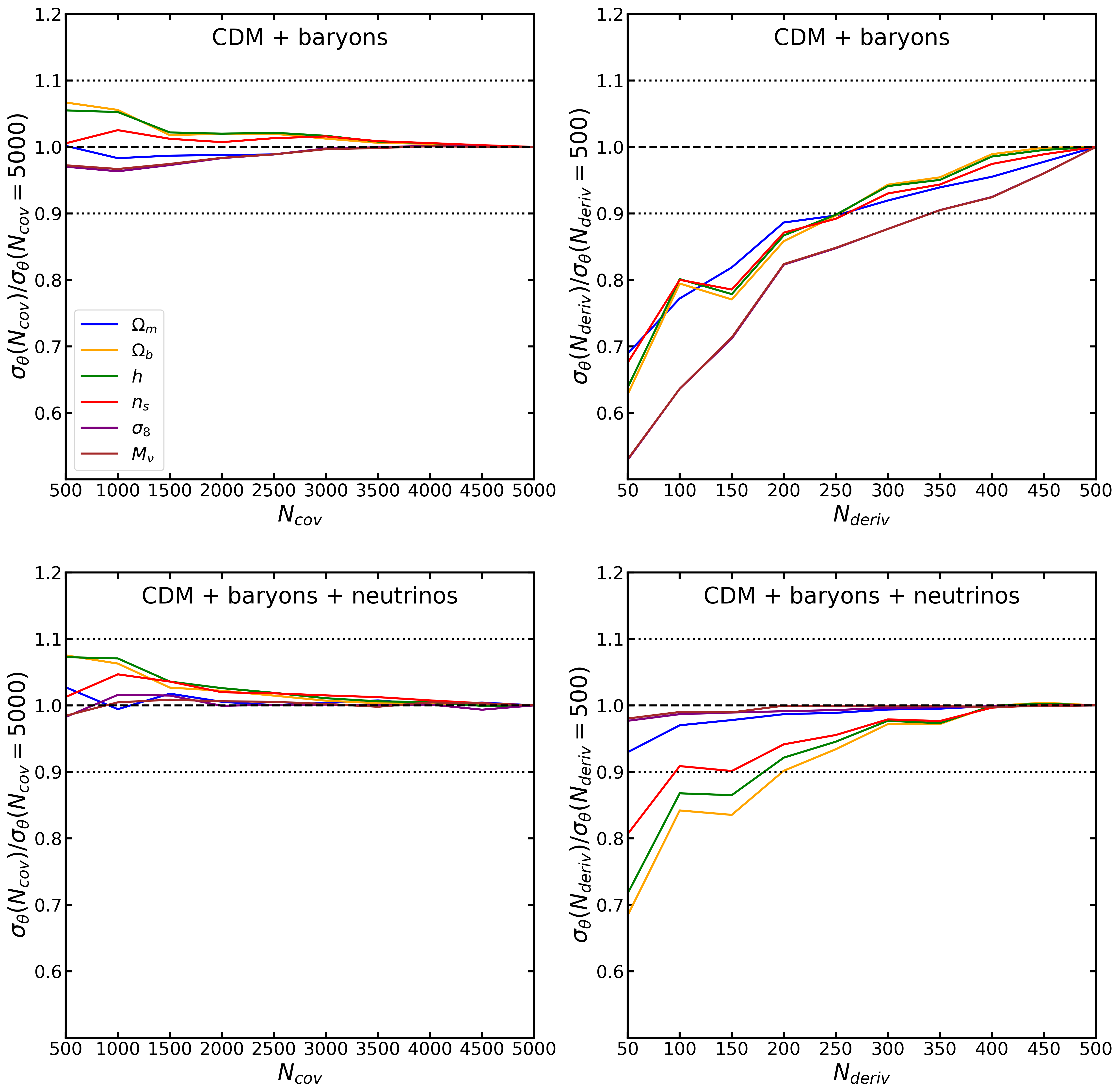}
	\caption{\label{fig:conver} Convergence of the marginalized errors from the combination of the MFs ($R_G=5 h^{-1}\mathrm{Mpc}$) and power spectrum on the cosmological parameters of $\Omega_{m}$, $\Omega_{b}$, $h$, $n_s$, $\sigma_{8}$ and $M_{\nu}$. The first column shows $\sigma_{\theta}(N_{cov})/\sigma_{\theta}(N_{cov}=5000)$, the ratio of Fisher forecasts obtained with covariance matrices estimated from $N_{cov}$ realizations to those obtained with covariance matrices estimated from $N_{cov}=5000$ realizations. Both forecasts are based on derivatives estimated with $N_{deriv}=500$ realizations. The second column shows $\sigma_{\theta}(N_{deriv})/\sigma_{\theta}(N_{deriv}=500)$, the ratio of Fisher forecasts obtained with derivatives estimated from $N_{deriv}$ realizations to those obtained with derivatives estimated from $N_{deriv}=500$ realizations. Both forecasts are based on covariance matrices estimated with $N_{cov}=5000$ realizations. The first row shows results from the combination of power spectrum and the MFs of the CDM + baryons field, while the second row shows results for the combination of power spectrum and the MFs of the CDM + baryons + neutrinos field. } 
\end{figure}

As mentioned in section~\ref{sec:fisher}, when the cosmological parameter space has a high dimension, the degeneracies between parameters lead to a Fisher matrix with a large condition number. The parameter covariance matrix is obtained by inverting the Fisher matrix, thus small instabilities in the Fisher matrix may result in much larger deviations in the parameter covariance matrix. In Figure~\ref{fig:conver}, we check how the marginalized errors from the combination of the power spectrum and the MFs vary when the number of realizations used to estimate the derivatives with respect to cosmological parameters $N_{\rm deriv}$ or the covariance matrix $N_{\rm cov}$ is increased. The maximum number of observables used in our work is only 160, hence $\sigma_{\theta}$ converges (varies $\lesssim 5\%$) when $N_{\rm cov}>1500$ for all parameters, for both the `cb' field and `m' field. The convergence of $\sigma_{\theta}$ with respect to $N_{\rm deriv}$ depends more on the level of degeneracies among parameters. For the `m' density field, degeneracy between $\sigma_8$ and $M_{\nu}$ is broken, thus $\sigma_{\theta}$ varies $\lesssim 3\%$ for $N_{\rm deriv}>300$, which is well converged. For the `cb' density field, $\sigma_{\theta}$ varies $\lesssim 7\%$ for $N_{\rm deriv}>400$, which is good convergence too, though due to limitation on the number of simulations used to calculate the derivatives, we cannot extend the curves to $N_{\rm deriv}>500$.

\bibliographystyle{JHEP}
\bibliography{reference}
\end{document}